\documentclass[11pt]{article}
\usepackage{axodraw}
\usepackage{epsfig}
\newcommand{\la}[1]{\label{#1}}
\newcommand{\be}{\begin{equation}}
\newcommand{\ee}{\end{equation}}
\newcommand{\ba}{\begin{eqnarray}}
\newcommand{\ea}{\end{eqnarray}}
\newcommand{\rmi}[1]{{\mbox{\scriptsize #1}}}
\newcommand{\fig}{Fig.~}
\newcommand{\eq}{Eq.~}
\newcommand{\se}{Sec.~}
\newcommand{\eqs}{Eqs.~}
\newcommand{\nr}[1]{(\ref{#1})}
\newcommand{\tr}{{\rm Tr\,}}
\newcommand{\nn}{\nonumber \\}
\newcommand{\fr}[2]{{\frac{#1}{#2}\,}}

\newcommand{\msbar}{{\overline{\mbox{\rm MS}}}}

\renewcommand{\vec}[1]{{\bf #1}}

\newcommand{\Ms}{M_{\rm s}}

\newcommand{\Nc}{N_{\rm c}}
\newcommand{\Tc}{T_{\rm c}}

\def\lsi{\raise0.3ex\hbox{$<$\kern-0.75em\raise-1.1ex\hbox{$\sim$}}}
\def\gsi{\raise0.3ex\hbox{$>$\kern-0.75em\raise-1.1ex\hbox{$\sim$}}}
\newcommand{\lsim}{\mathop{\;\lsi\;}}
\newcommand{\gsim}{\mathop{\;\gsi\;}}

\newcommand{\nF}[1]{n_\rmi{F{#1}}}
\newcommand{\nB}[1]{n_\rmi{B{#1}}}
\newcommand{\rmii}[1]{{\mbox{\tiny\rm{#1}}}}
\newcommand{\re}{\mathop{\mbox{Re}}}
\newcommand{\im}{\mathop{\mbox{Im}}}
\newcommand{\bsl}[1]{\,\slash\!\!\!\!{#1}\,}
\newcommand{\msl}[1]{\,\slash\!\!\!{#1}\,}
\newcommand{\Tint}[1]{{\hbox{$\sum$}\!\!\!\!\!\!\int}_{\!\!\!\!\raise-0.9ex\hbox{$\scriptstyle{#1}$}}}

\newcommand{\pic}[1]{\;\parbox[c]{30pt}{\begin{picture}(30,30)(0,0)
\SetWidth{1.0}\SetScale{1.0} #1 \end{picture}}\;}

\newcommand{\picc}[1]{\;\parbox[c]{60pt}{\begin{picture}(60,30)(0,0)
\SetWidth{1.0}\SetScale{1.0} #1 \end{picture}}\;}

\newcommand{\piccc}[1]{\;\parbox[c]{90pt}{\begin{picture}(90,30)(0,0)
\SetWidth{1.0}\SetScale{1.0} #1 \end{picture}}\;}

\def\Lwidth{1}

\def\Agl(#1,#2)(#3,#4,#5){\PhotonArc(#1,#2)(#3,#4,#5){\Lwidth}
{6.283 #3 mul 360 div #4 #5 sub #4 #5 sub mul sqrt mul Ldensity mul}}
\def\Lgl(#1,#2)(#3,#4){\Photon(#1,#2)(#3,#4){\Lwidth}
{#1 #3 sub #1 #3 sub mul #2 #4 sub #2 #4 sub mul add sqrt Ldensity mul}}
\def\Agh(#1,#2)(#3,#4,#5){\DashArrowArc(#1,#2)(#3,#4,#5){1}}
\def\Aagh(#1,#2)(#3,#4,#5){\DashArrowArcn(#1,#2)(#3,#5,#4){1}}
\def\Lgh(#1,#2)(#3,#4){\DashArrowLine(#1,#2)(#3,#4){1}}
\def\Lagh(#1,#2)(#3,#4){\DashArrowLine(#3,#4)(#1,#2){1}}
\def\Ahh(#1,#2)(#3,#4,#5){\DashCArc(#1,#2)(#3,#4,#5){1}}
\def\Lhh(#1,#2)(#3,#4){\DashLine(#1,#2)(#3,#4){1}}
\def\Aqu(#1,#2)(#3,#4,#5){\ArrowArc(#1,#2)(#3,#4,#5)}
\def\Aaqu(#1,#2)(#3,#4,#5){\ArrowArcn(#1,#2)(#3,#5,#4)}
\def\Lqu(#1,#2)(#3,#4){\ArrowLine(#1,#2)(#3,#4)}
\def\Laqu(#1,#2)(#3,#4){\ArrowLine(#3,#4)(#1,#2)}
\def\Aqq(#1,#2)(#3,#4,#5){\CArc(#1,#2)(#3,#4,#5)}
\def\Lqq(#1,#2)(#3,#4){\Line(#1,#2)(#3,#4)}
\def\Asc(#1,#2)(#3,#4,#5){\CArc(#1,#2)(#3,#4,#5)}
\def\Lsc(#1,#2)(#3,#4){\Line(#1,#2)(#3,#4)}

\def\scfc{0.7}  
\def\phgt{21}   
\def\pwc{21}    
\def\pwcc{42} 
\newcommand{\PIC}[4]{\;\parbox[c]{#2 pt}{\begin{picture}(#2,#3)(0,0)
\SetWidth{1.0}\SetScale{#4} #1 \end{picture}}\;}
\renewcommand{\pic}[1]{\PIC{#1}{\pwc}{\phgt}{\scfc}}

\renewcommand{\picc}[1]{\PIC{#1}{\pwcc}{\phgt}{\scfc}}
\def\Textt(#1,#2,#3){\Text(#1,#2)[t]{{$\scriptstyle #3$}}}
\def\Textb(#1,#2,#3){\Text(#1,#2)[b]{{$\scriptstyle #3$}}}
\def\Textl(#1,#2,#3){\Text(#1,#2)[l]{{$\scriptstyle #3$}}}
\def\Textr(#1,#2,#3){\Text(#1,#2)[r]{{$\scriptstyle #3$}}}
\def\Texttl(#1,#2,#3){\Text(#1,#2)[tl]{{$\scriptstyle #3$}}}
\def\Textbl(#1,#2,#3){\Text(#1,#2)[bl]{{$\scriptstyle #3$}}}
\def\Texttr(#1,#2,#3){\Text(#1,#2)[tr]{{$\scriptstyle #3$}}}
\def\Textbr(#1,#2,#3){\Text(#1,#2)[br]{{$\scriptstyle #3$}}}
\def\Textst(#1,#2,#3){\Text(#1,#2)[t]{{$\scriptscriptstyle #3$}}}
\def\Textsb(#1,#2,#3){\Text(#1,#2)[b]{{$\scriptscriptstyle #3$}}}
\def\Textsl(#1,#2,#3){\Text(#1,#2)[l]{{$\scriptscriptstyle #3$}}}
\def\Textsr(#1,#2,#3){\Text(#1,#2)[r]{{$\scriptscriptstyle #3$}}}
\def\Textstl(#1,#2,#3){\Text(#1,#2)[tl]{{$\scriptscriptstyle #3$}}}
\def\Textsbl(#1,#2,#3){\Text(#1,#2)[bl]{{$\scriptscriptstyle #3$}}}
\def\Textstr(#1,#2,#3){\Text(#1,#2)[tr]{{$\scriptscriptstyle #3$}}}
\def\Textsbr(#1,#2,#3){\Text(#1,#2)[br]{{$\scriptscriptstyle #3$}}}
\def\Lwidth{1}

\def\TopoSBnew(#1,#2,#3){\piccc{#1(0,15)(15,15) #2(30,15)(15,0,180)%
 #3(30,15)(15,180,360) #1(45,15)(60,15)%
 \Textl(45,30,\tilde R)\Textl(45,0,\tilde Q+\tilde R)\Textb(3,18,\tilde Q)}}
\def\Bqu(#1,#2)(#3,#4,#5){\SetWidth{2.0}\ArrowArc(#1,#2)(#3,#4,#5)%
\SetWidth{1.0}}
\def\ScatA{\picc{%
 \Laqu(15,30)(30,15)%
 \Lqu(12,15)(30,15)%
 \Laqu(15,0)(30,15)%
 \Lagh(30,15)(45,15)%
 \Textsl(6,21,1)%
 \Textsl(3,10.5,2)%
 \Textsl(6,0,3)%
 \Textsr(40,10.5,Q)%
}}
\def\ScatB{\picc{%
 \Lqu(15,30)(30,15)%
 \Lqu(30,15)(45,30)%
 \Laqu(15,0)(30,15)%
 \Lagh(30,15)(45,0)%
 \Textsl(5,21,2)%
 \Textsr(38,21,1)%
 \Textsl(5,0,3)%
 \Textsr(40,0,Q)%
}}
\def\ScatC{\picc{%
 \Laqu(15,30)(30,15)%
 \Laqu(30,15)(45,30)%
 \Laqu(15,0)(30,15)%
 \Lagh(30,15)(45,0)%
 \Textsl(5,21,1)%
 \Textsr(38,21,2)%
 \Textsl(5,0,3)%
 \Textsr(40,0,Q)%
}}
\def\ScatD{\picc{%
 \Laqu(15,30)(30,15)%
 \Lqu(30,15)(45,30)%
 \Lqu(15,0)(30,15)%
 \Lagh(30,15)(45,0)%
 \Textsl(5,21,1)%
 \Textsr(38,21,3)%
 \Textsl(5,0,2)%
 \Textsr(40,0,Q)%
}}
\def\ScatE{\picc{%
 \Laqu(15,15)(30,15)%
 \Laqu(30,15)(45,30)%
 \Lagh(30,15)(45,15)%
 \Lqu(30,15)(45,0)%
 \Textsl(5,10.5,1)%
 \Textsr(38,21,2)%
 \Textsr(40,10.5,Q)%
 \Textsr(38,0,3)%
}}
\def\ScatF{\picc{%
 \Lqu(15,15)(30,15)%
 \Lqu(30,15)(45,30)%
 \Lagh(30,15)(45,15)%
 \Lqu(30,15)(45,0)%
 \Textsl(5,10.5,2)%
 \Textsr(38,21,1)%
 \Textsr(40,10.5,Q)%
 \Textsr(38,0,3)%
}}
\def\ScatG{\picc{%
 \Laqu(15,15)(30,15)%
 \Lqu(30,15)(45,30)%
 \Lagh(30,15)(45,15)%
 \Laqu(30,15)(45,0)%
 \Textsl(5,10.5,3)%
 \Textsr(38,21,1)%
 \Textsr(40,10.5,Q)%
 \Textsr(38,0,2)%
}}
\def\ScatH{\pic{%
 \Lqu(0,15)(15,30)%
 \Laqu(0,15)(15,20)%
 \Lagh(0,15)(15,10)%
 \Lqu(0,15)(15,0)%
 \Textsr(18,21,1)%
 \Textsr(18,14,2)%
 \Textsr(19,6.5,Q)%
 \Textsr(18,0,3)%
}}
\def\Scata{\pic{%
 \Lqu(0,15)(15,30)%
 \Lagh(0,15)(15,15)%
 \Lgl(0,15)(15,0)%
 \Textsr(18,21,1)%
 \Textsr(20,10.5,Q)%
 \Textsr(18,0,2)%
}}
\def\Scatb{\picc{%
 \Lgl(15,15)(30,15)%
 \Lqu(30,15)(45,30)%
 \Lagh(30,15)(45,15)%
 \Textsl(5,10.5,2)%
 \Textsr(38,21,1)%
 \Textsr(40,10.5,Q)%
}}
\def\Scatc{\picc{%
 \Laqu(15,15)(30,15)%
 \Lgl(30,15)(45,30)%
 \Lagh(30,15)(45,15)%
 \Textsl(5,10.5,1)%
 \Textsr(38,21,2)%
 \Textsr(40,10.5,Q)%
}}
\def\Scatd{\picc{%
 \Laqu(15,30)(30,15)%
 \Lgl(12,15)(30,15)%
 \Lagh(30,15)(45,15)%
 \Textsl(6,21,1)%
 \Textsl(3,10.5,2)%
 \Textsr(40,10.5,Q)%
}}

\makeatletter \@addtoreset{equation}{section} \makeatother
\renewcommand{\theequation}{\arabic{section}.\arabic{equation}}
\makeatletter
\renewcommand\section{\@startsection {section}{1}{\z@}%
                                   {-5.5ex \@plus -1ex \@minus -.2ex}
                                   {2.3ex \@plus.2ex}%
                                   {\normalfont\large\bfseries}}
\renewcommand\subsection{\@startsection{subsection}{2}{\z@}%
                                     {-3.25ex\@plus -1ex \@minus -.2ex}%
                                     {1.5ex \@plus .2ex}%
                                     {\normalfont\normalsize\bfseries}}
\renewcommand\thesection {\@arabic\c@section}
\renewcommand\thesubsection   {\thesection.\@arabic\c@subsection}
\renewcommand{\@seccntformat}[1]{%
\csname the#1\endcsname.\hspace{1.0em}}
\makeatother

\begin{document}

\begin{titlepage}
\begin{flushright}
hep-ph/0612182\\
\end{flushright}
\begin{centering}
\vfill

{\Large{\bf Lightest sterile neutrino abundance within the $\nu$MSM}}

\vspace{0.8cm}

Takehiko~Asaka$^\rmi{a}$,
Mikko~Laine$^\rmi{b}$, 
Mikhail~Shaposhnikov$^\rmi{a}$ 

\vspace{0.8cm}

$^\rmi{a}${\em
Institut de Th\'eorie des Ph\'enom\`enes Physiques, EPFL, 
CH-1015 Lausanne, Switzerland}


\vspace{0.3cm}

$^\rmi{b}${\em
Faculty of Physics, University of Bielefeld, 
D-33501 Bielefeld, Germany\\}


\vspace*{0.8cm}
 
\mbox{\bf Abstract}

\end{centering}

\vspace*{0.3cm}
 
\noindent 
We determine the abundance of the lightest (dark matter) sterile
neutrinos created in the Early Universe due to active-sterile neutrino
transitions from the thermal plasma. Our starting point is the
field-theoretic formula for the sterile neutrino production rate,
derived in our previous work~[JHEP 06(2006)053], which allows to
systematically incorporate all relevant effects, and also to analyse
various hadronic uncertainties. Our numerical results differ
moderately from previous computations in the literature, and lead to
an absolute upper bound on the mixing angles of the dark matter
sterile neutrino. Comparing this bound with existing astrophysical
X-ray constraints, we find that the Dodelson-Widrow scenario, which
proposes sterile neutrinos generated by active-sterile neutrino
transitions to be the sole source of dark matter, is only possible for
sterile neutrino masses lighter than 3.5 keV (6 keV if all hadronic
uncertainties are pushed in one direction and the most stringent X-ray
bounds are relaxed by a factor of two).  This upper bound may conflict
with a lower bound from structure formation, but a definitive
conclusion necessitates numerical simulations with the non-equilibrium
momentum distribution function that we derive. If other production
mechanisms are also operative, no upper bound on the sterile neutrino
mass can be established.
\vfill
\noindent
 

\vspace*{1cm}
 
\noindent
January 2007

\vfill

\end{titlepage}

%
\section{Introduction}
\la{se:intro}

In recent works \cite{Asaka:2005an,Asaka:2005pn} it has been
demonstrated that an extension of the Minimal Standard Model (MSM) by
three right-handed gauge-singlet fermions --- sterile neutrinos ---   
with masses smaller than the electroweak scale,  allows to address a number of
phenomena which cannot be explained in the framework of the MSM. 
Indeed there exists a parameter choice within this model, 
called the $\nu$MSM  in refs.~\cite{Asaka:2005an,Asaka:2005pn},  
which is consistent with
experimentally observed neutrino masses and mixings,  provides a
candidate for dark matter particles, and can explain the baryon
asymmetry of the Universe~\cite{Asaka:2005pn}, through a CP-violating
redistribution of the lepton number among active  and sterile
flavours~\cite{Akhmedov:1998qx},  followed by an anomalous conversion
of the lepton number  in active flavours into a baryon
number~\cite{Kuzmin:1985mm}. Adding to this model one neutral scalar
field allows to accommodate  inflation~\cite{Shaposhnikov:2006xi}.
Moreover various astrophysical problems may find their
explanations~\cite{astro}.

In the $\nu$MSM, the role of dark matter is played by the lightest
sterile neutrino, with a mass in the keV range. This dark matter
candidate  was proposed a while ago in ref.~\cite{Dodelson:1993je}
from  different considerations, and studied later on in a number of 
works~\cite{Shi:1998km}--\cite{Boyanovsky:2006it}. It was pointed out
in ref.~\cite{Dodelson:1993je} that the couplings of the dark matter
sterile neutrino to charged leptons and active neutrinos can be so
weak that it never equilibrates in the Early Universe. As has been
stressed  recently~\cite{Boyarsky:2005us}--\cite{als}, this means
that even if the exact values of the mixing angles between sterile
and active neutrinos were known,  the primordial abundance of the
dark matter sterile neutrinos cannot  be predicted. The reason is
trivial: the kinetic equation describing  sterile neutrino production
is a first order differential equation in time, and in order to solve
it one must specify the initial condition, on which the solution
depends significantly, because inverse processes are much slower than
the rate of Universe expansion.   In more physical terms, one has to
know the physics  beyond the $\nu$MSM to set the initial conditions; 
an example of a complete framework,  involving the inflaton, can be
found in ref.~\cite{Shaposhnikov:2006xi}.

Though a prediction of the primordial abundance of the dark
matter sterile neutrinos requires physics beyond the $\nu$MSM,  an
upper limit on the mixing angle $\theta$ between sterile  and active
neutrinos  as a function of the sterile neutrino mass $\Ms$, 
\be
 \sin^2(2\theta) \lsim f(\Ms)
 \;,
 \label{main}
\ee
can be established. Indeed, some number of sterile neutrinos are
certainly produced in the Early Universe through the mixing with
active neutrinos, and this amount must be smaller than the  abundance
of dark matter, known from observations. This upper bound can  be
derived with a small number of extra assumptions, which we  formulate
as follows:

\begin{itemize}

\item[(i)] 
The $\nu$MSM is a good effective theory below energies of 
a few GeV, so that there are no interactions beyond those
included in the $\nu$MSM Lagrangian.

\item[(ii)] 
The standard Big Bang scenario is valid starting from 
temperatures above a few GeV.

\item[(iii)] 
The charge asymmetries of the plasma 
(particularly the asymmetries in the total lepton number, 
and in the various lepton flavours)
are small,  i.e.\ at most within
a few orders of magnitude  of the observed baryon asymmetry,  at
temperatures below a few GeV.

\item[(iv)] 
The masses of the two heavier sterile neutrinos  are large enough so
that they decay above temperatures of a few GeV.

\end{itemize}

The assumptions (i-iii) are crucial. For example, if the inflaton is
light, its interactions may result in sterile neutrino production
at a low scale~\cite{Shaposhnikov:2006xi}, making (i) invalid. For very low
reheating temperatures after inflation, the assumption (ii) is not satisfied 
and the production of sterile neutrinos may be suppressed
\cite{Gelmini:2004ah}. As for point (iii), it is known from
ref.~\cite{Shi:1998km}  that the rate of sterile neutrino production
is in fact greatly boosted if relatively large lepton asymmetries 
(corresponding to chemical potentials $\mu/T \gsim 10^{-5}$) are
present. On the other hand, if the crucial assumptions (i--iii) 
are valid, the assumption (iv) can be relaxed: once the 
result is known, the decays of heavier sterile neutrinos can
be taken into account {\em a posteriori}~\cite{Asaka:2006ek}. 
We will return to the last point below. 

There are quite a number of computations of the function  $f(\Ms)$
already existing in the literature, starting from  similar
assumptions~\cite{Dodelson:1993je,Dolgov:2000ew, Abazajian:2001nj,
Abazajian:2002yz,Abazajian:2005gj}.  They are based, however, on
kinetic equations which were not derived rigorously and which in fact
differ from the expressions obtained from the first principles of
statistical mechanics and quantum field theory~\cite{als}.  Moreover,
as discussed in ref.~\cite{als},  neither the hadronic scattering
contributions to the rate of sterile  neutrino production,  nor the
uncertainties in the hadronic equation-of-state,  have been 
exhaustively analyzed  in refs.~\cite{Dodelson:1993je,Dolgov:2000ew, 
Abazajian:2001nj,Abazajian:2002yz,Abazajian:2005gj}.

The aim of the present paper is therefore to apply the general
formalism of ref.~\cite{als} to find the amount and spectrum of the
sterile neutrinos created in active-sterile transitions. Our results
differ from those that have previously  appeared in the literature,
even though it turns out that the order of magnitude remains the same.

Given the results of the theoretical computation, avoiding the
overclosure of the Universe allows us to establish the constraint in
\eq\nr{main}.  We can, however, also carry out a comparison with other
astrophysical and cosmological limits on the properties of sterile
neutrinos.  Indeed, the sterile neutrino radiative decays $N \to \nu
\gamma$ produce a feature in the diffuse X-ray
background~\cite{Dolgov:2000ew,Boyarsky:2005us} or a line in the X-ray
spectrum in the direction where dark matter is accumulated (such as
clusters of galaxies \cite{Abazajian:2001vt,Boyarsky:2006zi}, dwarf
galaxies \cite{Boyarsky:2006fg}, or galaxies
\cite{Abazajian:2001vt,Boyarsky:2006fg,Riemer-Sorensen:2006fh,Watson:2006qb}).
The position of this line, if found, determines the mass $\Ms$ of the
sterile neutrino, while the line intensity would fix the mixing angle
$\theta$.  No feature or line has been observed so far, which places a
constraint on the mixing angle of the form $\sin^2(2\theta) \lsim
f_X(\Ms)$ \cite{Boyarsky:2005us,Boyarsky:2006zi,
  Boyarsky:2006fg,Riemer-Sorensen:2006fh,Watson:2006qb,
  Boyarsky:2006kc,Boyarsky:2006ag,Riemer-Sorensen:2006pi,
  Abazajian:2006jc,Boyarsky:2006hr}.  An exclusion plot from
refs.~\cite{Boyarsky:2006fg,Boyarsky:2006ag} is reproduced in
\fig\ref{fig:exclusion} below, and can be compared with \eq\nr{main}.

Another, completely independent constraint comes from cosmological
structure formation, particularly in the form of Lyman-$\alpha$
forest  observations~\cite{Hansen:2001zv}--\cite{Viel:2006kd}.  Being
relatively light, dark matter sterile neutrinos would play the role
of {\em warm} dark matter, with a  free-streaming length exceeding
greatly that of cold dark matter.  This erases inhomogeneities on the
smallest scales.  An observation of the small scale structures puts,
therefore,  an upper bound on the free-streaming length and,
consequently, on the average velocity of the dark matter particles. 
This converts to a lower bound on the inverse velocity, 
which can be expressed roughly as 
$\Ms \langle|\vec{q}_\rmi{a}|\rangle / \langle |\vec{q}_\rmi{s}|
\rangle \gsim M_0$,
where $\langle|\vec{q}_\rmi{a}|\rangle$ 
and $\langle |\vec{q}_\rmi{s}| \rangle$
are the average momenta of active
and sterile neutrinos, respectively,  at the moment of structure
formation,  and the value of $M_0$ is as large as $M_0 \simeq 14.4$ keV
according to ref.~\cite{Seljak:2006qw}. The computations of 
refs.~\cite{Dodelson:1993je,Abazajian:2001nj,Abazajian:2005gj} lead
to a nearly-thermal spectrum of sterile neutrinos which is somewhat
shifted in the infrared in comparison with the Fermi-Dirac
distribution. According to ref.~\cite{Abazajian:2005gj}, 
$\langle|\vec{q}_\rmi{s}|\rangle / \langle |\vec{q}_\rmi{a}|
\rangle \simeq 0.9$
(uncertainties in this number and its dependence on the sterile
neutrino mass will be discussed below). This leads to 
$\Ms > 13$ keV \cite{Seljak:2006qw},  
to again be compared with \eq\nr{main}.
Note that ref.~\cite{Viel:2006kd} gives a somewhat weaker bound $\Ms \gsim
10$ keV.  We stress that
these mass bounds do depend on the mechanism of sterile neutrino
production through the momentum distribution function of sterile 
neutrinos~\cite{Asaka:2006ek}. A model-independent constraint 
(the Tremaine-Gunn bound) on the mass  
coming from the analysis of the rotational curves of
dwarf satellite galaxies \cite{Tremaine:1979we,Lin:1983vq} is in fact
much weaker and reads $\Ms \gsim 0.3$ keV \cite{Dalcanton:2000hn}.

These three different types of constraints, Eq. (\ref{main}), X-ray,
and Lyman-$\alpha$ or the Tremaine-Gunn bound, determine the allowed values 
for the mass and mixing angles of the dark matter sterile neutrino, necessary 
for planning for their search in space missions~\cite{Boyarsky:2006hr}
and in laboratory
experiments \cite{Bezrukov:2006cy}. In addition, the validity of the 
Dodelson-Widrow scenario, where only thermal production is taken into
account (so that \eq\nr{main} becomes an equality), can be tested.

The plan of the paper is the following.  We start by reviewing the
formalism  of ref.~\cite{als} in~\se\ref{se:general}. In
\se\ref{se:leptonic} we apply it to the leptonic contribution to the
sterile neutrino production rate, while in \se\ref{se:kinetic} the
kinetic equation for the sterile neutrino abundance and its solution in
the cosmological context are discussed.  In \se\ref{se:qcd} we
analyse the hadronic contributions that play  a role in the solution
of the kinetic equation, and their uncertainties. In
\se\ref{se:numerics} we combine all these results together  and
present numerical estimates for the relic sterile neutrino abundance 
as a function of $\Ms$ and $\theta$.  In \se\ref{se:DW} we compare our
results with the observational constraints mentioned above,  and
discuss the viability of the Dodelson-Widrow scenario for sterile
neutrino production. We conclude in \se\ref{se:concl}. In this paper
the notations of ref.~\cite{als} will be used unless stated otherwise.

%
\section{Review of the general formalism}
\la{se:general}
As has been discussed in ref.~\cite{als}, the assumptions (i--iv)
allow to derive, from first principles, a general formula  for the
sterile neutrino production rate, provided that we choose  parameter
values consistent with the observational constraints  mentioned
above. The restriction to temperatures below a few GeV in these
assumptions means that the electroweak symmetry is broken, whereby 
it is an excellent approximation to replace the Higgs field by its 
vacuum expectation value $v\simeq 246$~GeV. Moreover combining the
two sets of observational  constraints restricts the mixing
parameter(s) $\theta$  to be small. To be precise, there are several
mixing parameters, $\theta_{\alpha I}$,  where $I$ is the sterile
neutrino flavour and $\alpha$ is the active neutrino flavour. We
define  $\theta_{\alpha I}^2 \equiv |M_D|_{\alpha I}^2/M_I^2$, where
$|M_D|_{\alpha I} \equiv |v F_{\alpha I}| /\sqrt{2}$;  $F_{\alpha I}$
are the neutrino Yukawa couplings;  and $M_I$ are the sterile
neutrino Majorana masses.  In the following we choose the lightest
sterile neutrino to  correspond to $I \equiv 1$. Because of the
smallness of $\theta_{\alpha I}$, $\theta_{\alpha I} \lsim 10^{-3}$
(or, in terms of Yukawa couplings, $|h_{\alpha 1}| \lsim 10^{-11}$, 
$|h_{\alpha 2}|, |h_{\alpha 3}| \lsim 10^{-7}$), it is perfectly
sufficient to restrict to leading order in a Taylor  series in
$\theta_{\alpha I}^2$, which simplification plays an  essential role
in the first-principles derivation presented in  ref.~\cite{als}.
(For $I=2,3$ these constraints derive from 
the baryon asymmetry of the Universe~\cite{Asaka:2005pn}.) 
For future reference, let us also define
\be
 \theta^2 \equiv \sum_{\alpha = e,\mu,\tau} \theta_{\alpha 1}^2 
 \;; \la{thetatotal}
\ee
this angle corresponds to that in \eq\nr{main}.

With these premises, the phase space density $n_I(t,\vec{q})$ of
sterile neutrinos in either spin state, 
\be
 n_I(t,\vec{q})= \sum_{s=1,2}
 \frac{{\rm d}
 N_I^{(s)}(t,\vec{x},\vec{q})}{{\rm d}^3 \vec{x}\,{\rm d}^3 \vec{q}} 
 \;,
\ee
obeys the equation 
\be
  \biggl( \frac{\partial}{\partial t} - 
  H q_i \frac{\partial}{\partial q_i}\biggr) 
  n_I(t,\vec{q}) 
  = R(T,\vec{q})
 \;, 
 \la{expansion}
 \la{kinetic}
\ee
where $H$ is the Hubble parameter,  $H={\rm d}\ln a(t)/ {\rm d}t$, 
and $q_i$ are the spatial components of $\vec{q}$. Note that in
thermal equilibrium, $n_I(t,q)=2 \nF{}(q^0)/(2\pi)^3$,  where $\nF{}$
is the Fermi distribution function.  The source term reads~\cite{als}
\ba
 && \hspace*{-1cm}
 R(T,\vec{q}) 
  =  \frac{4 \nF{}(q^0)}{(2\pi)^3 2 q^0}
   \sum_{\alpha = 1}^{3}  
   \frac{|M_D|^2_{\alpha I}}
   {\{[Q + \re\Sigma]^2 - [\im \Sigma]^2 \}^2 + 
       4 \{[Q + \re\Sigma]\cdot \im\Sigma \}^2 } \times
 \la{master2} \\ 
 &&  \times   
 \tr \Bigl\{
  \bsl{Q} a_L \Bigl(
    2 [Q + \re\Sigma]\cdot\im\Sigma 
      \;\, [ \bsl{Q} + \re \bsl{\Sigma} ]
    - \{
         [Q + \re \Sigma]^2 - [\im\Sigma]^2 
      \} \, \im\bsl{\Sigma} 
  \Bigr) a_R
  \Bigr\}
  \;, \nonumber
\ea 
where   $Q$ is the on-shell four-momentum of the sterile neutrino
(i.e.\ $Q^2 = M_I^2$);  $\Sigma \equiv \Sigma_{\alpha\alpha}(Q)$ is
the self-energy of the active neutrino of flavour $\alpha$;  and
$a_L, a_R$ are the chiral projectors. It is obvious that 
$n_I(t,\vec{q})$ and $R(T,\vec{q})$ are functions of $q \equiv
|\vec{q}|$  only, and we will use the corresponding simplified
notation in the following.  

Now, the real part $\re\Sigma$  of the active neutrino self-energy is
generated at 1-loop level through  $W$ and $Z$-boson exchange. At low
energies, the result can be expanded  in $1/m_W^2$, $1/m_Z^2$. In the
absence of leptonic chemical potentials,  the first term in the
expansion vanishes, and the leading contribution comes from the second term.
Writing the result as 
\be
 \re \bsl{\Sigma}_{\alpha\alpha}(Q) = 
 \bsl{Q} a_{\alpha\alpha}(Q) + \msl{u} b_{\alpha\alpha}(Q)
 \;,  \la{Restruct} 
\ee
where $u=(1,\vec{0})$,  we note that the function
$a_{\alpha\alpha}(Q)$ can be ignored, since  it is small compared
with the tree-level term $\bsl{Q}$.  On the other hand the latter
structure in \eq\nr{Restruct} does not appear at tree-level, and
needs to be kept. For $q\ll m_W$ it reads~\cite{ReSigmaold,ReSigma}
\be
 b_{\alpha\alpha}(Q) = \frac{16 G_F^2}{\pi\alpha_w} q^0
 \Bigl[
   2 \phi(m_{l_\alpha}) + \cos^2\! \theta_\rmii{W}\, \phi(m_{\nu_\alpha})
 \Bigr]
 \;, \la{bQ}
\ee
where $G_F = g_w^2/4\sqrt{2} m_W^2$ is the Fermi constant,
$m_{l_\alpha}$ is the mass 
of the charged lepton of generation $\alpha$
($l_1 \equiv e, l_2 \equiv \mu, l_3 \equiv \tau$), and
$m_{\nu_\alpha} = 0$ is the mass of the MSM active neutrino.
The function $\phi$ is finite and easily evaluated numerically: 
\be
 \phi(m) = 
 \int \! \frac{{\rm d}^3 \vec{p}}{(2\pi)^3}
 \frac{\nF{}(E)}{2 E}
 \biggl[
   \fr43 |\vec{p}|^2 + m^2 
 \biggr]_{E = \sqrt{|\vec{p}|^2 + m^2}} 
 \;. \la{phim}
\ee
Note, in particular, that $\phi(0) = 7 \pi^2 T^4/360$. The functions
$b_{\alpha\alpha}$ are plotted  in \fig\ref{fig:baa}.

\begin{figure}[t]

\centerline{%
\epsfysize=9.0cm\epsfbox{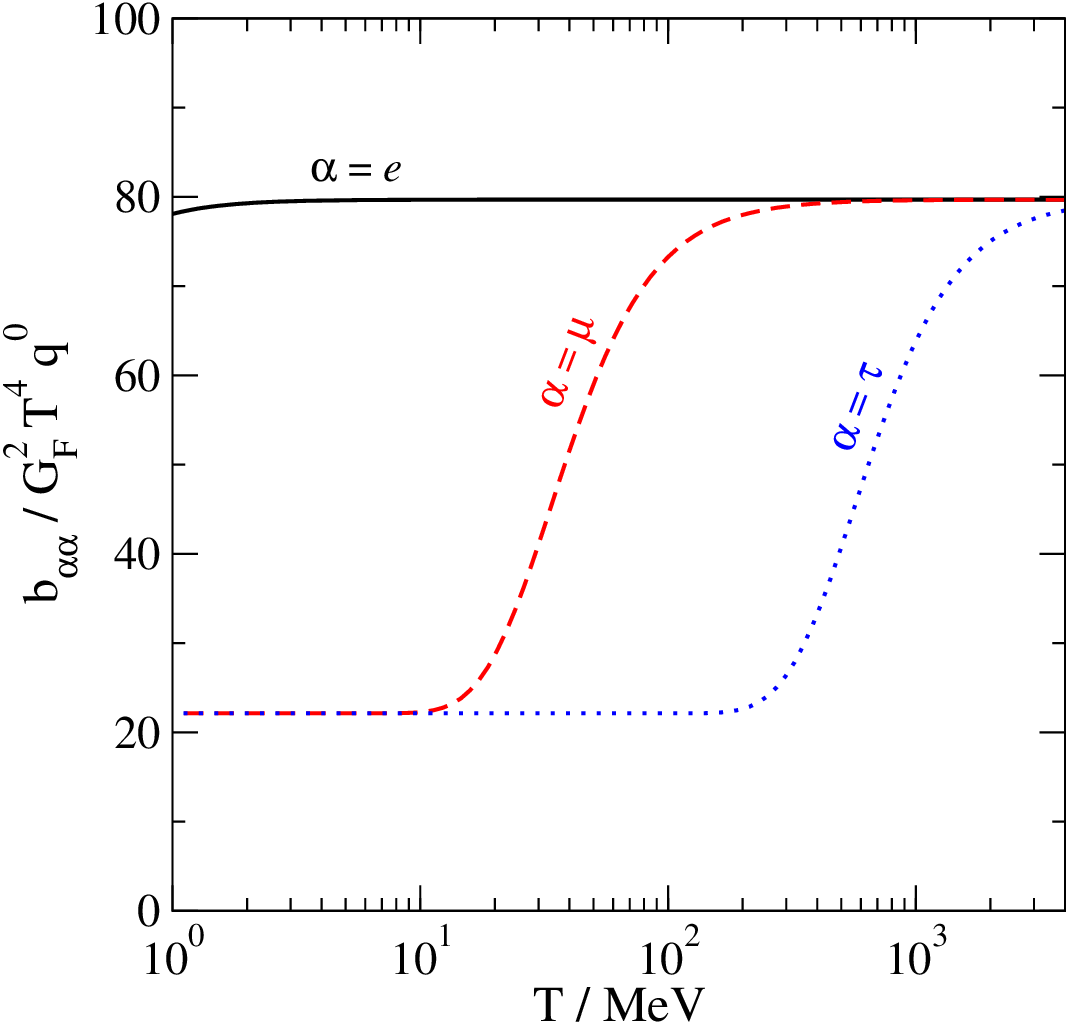}%
}

\caption[a]{\small
The function $b_{\alpha\alpha}(Q)$ that determines 
the real part $\re\Sigma_{\alpha\alpha}(Q)$ 
(cf.\ \eq\nr{Restruct}), in units
of $G_F^2 T^4 q^0$, as a function of the temperature $T$ 
and the active neutrino flavour $\alpha$, 
with $\alpha = e, \mu, \tau$.
We have assumed here that $|q^0| \ll m_W$.
} 

\la{fig:baa}
\end{figure}

As concerns the imaginary part $\im\Sigma$, there is a 1-loop 
contribution from the same graphs as for $\re\Sigma$,  but it is
exponentially suppressed, $\sim \exp(-m_W/T)$ (cf.\
\se\ref{se:1loop}). Therefore, at low temperatures,  the dominant
contribution is generated by those 2-loop graphs which are not
exponentially suppressed.  The dependence on $Q$ is more complicated
than in \eq\nr{bQ}:  in general $\im\Sigma(Q) =  G_F^2 T^5 f(Q,T)$, 
where $f$ is a non-trivial  dimensionless function,  which is
numerically of order unity. 

Given that $\im\Sigma \ll \re\Sigma$ because  of the
$\alpha_w$-suppression, and that  $\re\Sigma \approx u\,
b_{\alpha\alpha}(Q)$ because of the reasons mentioned above,  the
expression in \eq\nr{master2} can be simplified. Carrying out the
Dirac traces;  combining equivalent terms; and re-introducing Dirac
traces  and chiral projectors $a_L,a_R$ around $\im\bsl{\Sigma}$ at
the end,  as a reminder of our convention of showing them
explicitly~\cite{als},  we arrive at
\ba
 && \hspace*{-1cm}
 R(T,\vec{q}) 
  \approx  \frac{4 \nF{}(q^0)}{(2\pi)^3 2 q^0}
   \sum_{\alpha = 1}^{3}  
   {|M_D|^2_{\alpha I}}
   \times
   \nn  
 &&  \times   
   \frac{ 
        (M_I^2 - b^2)\,
        \tr \Bigl\{ \bsl{Q} a_L \im\bsl{\Sigma} a_R \Bigr\} + 
        2 (M_I^2 + q^0 b )\, b \, 
        \tr \Bigl\{ \msl{u} a_L \im\bsl{\Sigma} a_R \Bigr\} 
   }{\Bigl\{ M_I^2 + 2 q^0 b + b^2 \Bigr\}^2}
  \;, \la{master3} 
\ea 
where  $b \equiv b_{\alpha\alpha}(Q)$, and we made use of $Q^2 = M_I^2$.

Now, let us estimate the orders of magnitude that  are relevant for
\eq\nr{master3} (for analogous earlier  discussions see, e.g.,
refs.~\cite{Dodelson:1993je,Dolgov:2000ew}). We consider momenta of
order $|\vec{q}|\sim T$,  where the rate turns out to peak;  
temperatures in the range 1~MeV $\lsim T \lsim$ 10 GeV; and sterile
neutrino masses in the range  $10^{-1}$~keV $\lsim M_1 \lsim 10^3$
keV. Thereby $q^0 \sim T \gg M_1$, and $b \sim 50 G_F^2 T^5$ (cf.\
\fig\ref{fig:baa}). We now identify two different temperature
scales:  
\begin{itemize}

\item
We define $T_{1}$ such that $M_1^2 \sim 2 q^0 b$ for $T\sim T_{1}$, i.e.
\be
 T_{1} \sim \biggl( \frac{M_1}{10\, G_F} \biggr)^{\fr13} 
     \sim 200\, \mbox{MeV}\,
     \biggl( \frac{M_1}{\mbox{keV}} \biggr)^{\fr13} 
 \;.
\ee
Note that for $T \sim T_{1}$, $b^2\ll M_I^2$, and obviously also 
$b \ll q^0$. 

\item
We define $T_{2}$ such that $M_1^2 \sim b^2$ for $T\sim T_{2}$, i.e.
\be
 T_{2} \sim \biggl( \frac{M_1}{50\, G_F^2} \biggr)^{\fr15}
     \sim 2.7\, \mbox{GeV}\,
     \biggl( \frac{M_1}{\mbox{keV}} \biggr)^{\fr15} 
 \;.
\ee
Note that for $T\sim T_{2}$, $q^0 b \gg M_1^2$, while it is 
still true that $b \ll q^0$.

\end{itemize}
Let us then estimate the magnitude of the second row 
in \eq\nr{master3} for various temperatures. For simplicity
we set $\im\Sigma \sim G_F^2 T^5$, since the precise magnitude 
plays no role, given that $\im\Sigma$ appears linearly in 
all terms in the numerator. We observe that: 
\begin{itemize}

\item
For $T \ll T_{1}$, $M_1^2$ dominates in magnitude over all thermally
generated terms. The second row 
of \eq\nr{master3} then evaluates to $\sim G_F^2 T^6/M_1^2$, 
i.e. decreases fast at low $T$.

\item
For $T \sim T_{1}$, the second row evaluates to $\sim 1/10^2$. 
This is the peak value.

\item
For $T \gg T_{1}$, the denominator is dominated by $(2 q^0 b)^2$
and thus starts to increase.  
The second row of \eq\nr{master3} 
then evaluates to $\sim M_1^2/10^4 G_F^2 T^6$, 
i.e. decreases again fast.

\item
Once $T \sim T_{2}$, the two terms in the numerator of \eq\nr{master3}
are of the same order of magnitude. In fact, the first term becomes
negative, but this effect is compensated for by the second term, so 
that the expression as a whole remains positive. 
The overall magnitude is now $\sim G_F^2 T_{2}^4$, 
i.e.\ about $1/10^6$ of that at $T \sim T_{1}$.

\item
For $T \gg T_{2}$, the second row increases as $\sim G_F^2 T^4$.
Given that $G_F \approx 1/(290\, \mbox{GeV})^2$, however, these 
contributions remain very small in the region $T \lsim$~10~GeV. 
Moreover in practice our rough
approximation turns out to be an overestimate, when compared with the
exact numerical evaluation. 

\end{itemize} 

To confirm this qualitative picture,  we have verified numerically
that if one starts  the integration (to be specified in
\se\ref{se:kinetic}) from a high temperature  $T \sim 10\, \mbox{GeV}
\gg T_{2}$,  then both terms in the numerator of \eq\nr{master3} need
to be taken  into account, in order to obtain a positive production
rate (in the range $T \gsim (5 ... 10)$ GeV the exponentially 
suppressed 1-loop contribution to $\im\Sigma$ 
needs to be taken account as well, 
but this does not change the conclusions). 
Their combined contribution from the  high-temperature region
is, however, completely negligible. For the dark matter sterile neutrino
it is therefore   in practice enough to start the integration  from
$T \sim T_{2}\sim$ a few GeV, and keep only the first term in the 
numerator, which is what we will do in the following. 

%
\section{Leptonic contribution to the production rate}
\la{se:leptonic}

%
\subsection{1-loop effects}
\la{se:1loop}

We now proceed to determine $\im\Sigma$.
We assume that the temperature is well below
the temperature $T_\rmi{ew}$ where the electroweak crossover 
takes place, $T_\rmi{ew} \sim 100 - 200$~GeV. 
In this situation the Higgs phenomenon provides 
for a good tool  for carrying out perturbative computations.  
The simplest contributions to $\im\Sigma_{\alpha\alpha}$ come from
the same 1-loop graphs as were considered for 
$\re\Sigma_{\alpha\alpha}$, involving a single  $W^\pm$ or
$Z^0$-propagator; however we do not carry out any expansion
in the inverse $W^\pm$ or $Z^0$ masses like for $\re\Sigma_{\alpha\alpha}$.  
A straightforward computation in Feynman gauge yields
\ba
 & &  \hspace*{-3.5cm}
 \im\bsl{\Sigma}_\rmi{lep}^\rmi{1-loop}(Q) = 
 \pi\alpha_w \nF{}^{-1}(q^0)  \sum_{C = W,Z} p_C 
 \int \! \frac{{\rm d}^3 \vec{p}_1}{(2\pi)^3 2 E_1} \,
 \int \! \frac{{\rm d}^3 \vec{p}_2}{(2\pi)^3 2 E_2} \,
 \bsl{P}_1
 \times  
 \nn 
 & \times \biggl\{ & \!\!\!
 (2\pi)^4 \delta^{(4)}(P_1+P_2-Q) \, 
 \nF{1}\nB{2} 
 + \Scatd \nn & + & \!\!\!
 (2\pi)^4 \delta^{(4)}(P_2-P_1-Q) \,
 \nB{2}(1-\nF{1})
 + \Scatb \nn & + & \!\!\!
 (2\pi)^4 \delta^{(4)}(P_1-P_2-Q) \, 
 \nF{1}(1+\nB{2})
 + \Scatc \nn & + & \!\!\!
 (2\pi)^4 \delta^{(4)}(P_1+P_2+Q )\, 
 (1-\nF{1})(1+\nB{2})
 \biggr\}
 , \Scata  \nn 
 \la{pert1}
\ea
where $\nF{i} \equiv \nF{}(E_i)$, $\nB{i} \equiv \nB{}(E_i)$; and
$p_W \equiv 2$, $p_Z \equiv \cos^{-2}\theta_\rmii{W} $ are the ``weights''
of the charged and neutral current channels. 
Furthermore, $P_i \equiv (E_i,\vec{p}_i)$ are on-shell four-momenta, 
\be
 E_1 \equiv 
 \sqrt{\vec{p}_1^2 + m_{l_C}^2}
 \;, \quad
 E_2 \equiv \sqrt{\vec{p}_2^2 + m_C^2}
 \;, \la{pre_Es}
\ee
where $m_{l_W} \equiv m_{l_\alpha}$ 
and $m_{l_Z} \equiv m_{\nu_\alpha} = 0$.
The graphs in \eq\nr{pertB} illustrate the various processes, 
with the wavy line  indicating the weak gauge boson. 
The phase space integrals remaining 
can be evaluated numerically as explained in Appendix~A.

For the masses that we are interested in, 
$m_{l_\alpha}, M_I \ll m_W$, only one of the channels in
\eq\nr{pert1} gives a non-zero contribution, namely the 
second one.

Now, as discussed in the previous section, the dominant production of $N_I$
takes place at temperatures much below $m_W$, 
say, $T \lsim m_W / 10$. 
In this case the phase space factor $\nB{2}$ guarantees that the 1-loop
contribution is suppressed by at least $\sim \exp(-m_W/T)$, and thus
vanishingly small. 
One might worry that thermal effects on the $W^\pm$ boson mass
make $m_W(T)$ smaller than naively expected; at the very small
temperatures we are intested in, however, this effect is negligible. 
Therefore, the 1-loop contribution is indeed inessential for our 
purposes, and does not play a role in the following. 

%
\subsection{2-loop effects}
\la{se:2loop}

We now move to specifying the 2-loop  contribution to $\im\Sigma$  in
\eq\nr{master3}, originating from intermediate states containing
leptons only. These effects can be reliably treated within
perturbation theory, and are not exponentially suppressed at low
temperatures. At the same time, previous evaluations in the
literature have made used of  phenomenological approximations which
are not part of the strict  perturbative computation, particularly in
order to simplify the  Dirac structures that enter the sterile
neutrino production rate.  In the following we evaluate the leptonic
contributions without any such approximations. 

In order to proceed, it is actually helpful to  start by recalling
the contribution that emerges from  a pair of free quarks, with
masses $m_2,m_3$; the derivation of this result  has been discussed
in explicit detail in ref.~\cite{als}. The final result,  given by
\eq(3.41) of ref.~\cite{als}, reads:
\ba
 & &  \hspace*{-1cm}
 \im\bsl{\Sigma}_\rmi{had}^\rmi{2-loop}(Q) = 
 2 \Nc G_F^2 \nF{}^{-1}(q^0)  \sum_{C = W,Z} p_C 
 \int \! \frac{{\rm d}^3 \vec{p}_1}{(2\pi)^3 2 E_1} \,
 \int \! \frac{{\rm d}^3 \vec{p}_2}{(2\pi)^3 2 E_2} \,
 \int \! \frac{{\rm d}^3 \vec{p}_3}{(2\pi)^3 2 E_3} \,
 \times  
 \nn 
 & \times \biggl\{ & \!\!\!
 (2\pi)^4 \delta^{(4)}(P_1+P_2+P_3-Q )\, 
 \nF{1}\nF{2}\nF{3} \,
 \mathcal{A}(-m_{l_C},m_2,-m_3)
 + \ScatA \nn & + & \!\!\!
 (2\pi)^4 \delta^{(4)}(P_2+P_3 -P_1-Q) \,
 \nF{2}\nF{3}(1-\nF{1}) \,
 \mathcal{A}(m_{l_C},m_2,-m_3)
 + \ScatB \nn & + & \!\!\!
 (2\pi)^4 \delta^{(4)}(P_1+P_3-P_2-Q) \, 
 \nF{1}\nF{3}(1-\nF{2}) \,
 \mathcal{A}(-m_{l_C},-m_2,-m_3)
 + \ScatC \nn & + & \!\!\!
 (2\pi)^4 \delta^{(4)}(P_1+P_2-P_3-Q) \, 
 \nF{1}\nF{2}(1- \nF{3}) \,
 \mathcal{A}(-m_{l_C},m_2,m_3)
 + \ScatD \nn & + & \!\!\!
 (2\pi)^4 \delta^{(4)}(P_1-P_2-P_3-Q) \,
 \nF{1}(1-\nF{2})(1-\nF{3}) \,
 \mathcal{A}(-m_{l_C},-m_2,m_3)
 + \ScatE \nn & + & \!\!\!
 (2\pi)^4 \delta^{(4)}(P_2-P_1-P_3-Q) \, 
 \nF{2}(1-\nF{1})(1-\nF{3}) \,
 \mathcal{A}(m_{l_C},m_2,m_3)
 + \ScatF \nn & + & \!\!\!
 (2\pi)^4 \delta^{(4)}(P_3 -P_1-P_2-Q) \,
 \nF{3} (1-\nF{1})(1-\nF{2}) \,
 \mathcal{A}(m_{l_C},-m_2,-m_3)
 + \ScatG \nn & + & \!\!\!
 (2\pi)^4 \delta^{(4)}(-P_1-P_2-P_3-Q ) \, 
 (1-\nF{1})(1-\nF{2})(1-\nF{3}) \,
 \mathcal{A}(m_{l_C},-m_2,m_3)
 \biggr\}
 , \ScatH  \nn 
 \la{pertB}
\ea
where $\nF{i} \equiv \nF{}(E_i)$;
\be
 \mathcal{A}(m_{l_C},m_2,m_3) \equiv
 \gamma^{\mu} (\bsl{P}_{\! 1} + m_{l_C}) \gamma^\nu
 \, \tr\Bigl[
  (\bsl{P}_{\! 2} + m_2) \gamma_\mu \Gamma 
  (\bsl{P}_{\! 3} + m_3) \gamma_\nu \Gamma 
 \Bigr]
 \;;
\ee
$p_W \equiv 2$, $p_Z \equiv 1/2$ are the ``weights'' of the charged
and neutral current channels; and  $m_{l_C}$ has the same meaning as
in \eq\nr{pre_Es}. Furthermore, $P_i \equiv (E_i,\vec{p}_i)$ are
again on-shell  four-momenta, now with the mass assignments
\be
 E_1 \equiv 
 \sqrt{\vec{p}_1^2 + m_{l_C}^2}
 \;, \quad
 E_2 \equiv \sqrt{\vec{p}_2^2 + m_2^2}
 \;, \quad
 E_3 \equiv \sqrt{\vec{p}_3^2 + m_3^2}
 \;. \la{Es}
\ee
The graphs in \eq\nr{pertB} illustrate the various processes. 
Further details (in particular the values of the masses $m_2,m_3$ and
of the Dirac matrix $\Gamma$)  will be explained presently.

Considering then the leptonic contributions, it is obvious that  the
same eight kinematic possibilities  will appear as in \eq\nr{pertB},
just with different masses and coefficients.  On the other hand
additional terms appear as well,  since Z-exchange can proceed
through  more channels than in the hadronic case. 

Given that the kinematic possibilities are identical, however,  it is
possible to ``factorise'' the result into a ``Dirac part'' and a
``kinematic part''. It is therefore enough to show the  result for
the Dirac part by considering one of the kinematic channels only; for
simplicity we choose the alternative  in \eq\nr{pertB} where  all
masses are positive inside the $\mathcal{A}$-function. A
straightforward computation parallelling the one in ref.~\cite{als}
then produces the result 
\ba
 & &  \hspace*{-1.5cm}
 \im\bsl{\Sigma}_\rmi{lep}^\rmi{2-loop}(Q) = 
 2 G_F^2 \nF{}^{-1}(q^0)  
 \int \! \frac{{\rm d}^3 \vec{p}_1}{(2\pi)^3 2 E_1} \,
 \int \! \frac{{\rm d}^3 \vec{p}_2}{(2\pi)^3 2 E_2} \,
 \int \! \frac{{\rm d}^3 \vec{p}_3}{(2\pi)^3 2 E_3} \,
 \times  
 \nonumber\\[1mm] 
 & \times \biggl\{ & \!\!\!
 2 \gamma^\mu  (\bsl{P}_{\! 1} + m_{l_\alpha}) \gamma^\nu
 \sum_{\beta = e,\mu,\tau}
 \tr\Bigl[ 
  (\bsl{P}_{\! 2} + m_{l_\beta})
  \gamma_\mu
  a_L 
  (\bsl{P}_{\! 3} + m_{\nu_\beta})
  \gamma_\nu
  a_L 
 \Bigr]\; + 
 \nonumber\\[1mm]
 & + & \!\!\!
 \fr12 \gamma^\mu  (\bsl{P}_{\! 1} + m_{\nu_\alpha}) \gamma^\nu
 \sum_{\beta  = e,\mu,\tau}
 \tr\Bigl[ 
  (\bsl{P}_{\! 2} + m_{l_\beta})
  \gamma_\mu
  \Gamma 
  (\bsl{P}_{\! 3} + m_{l_\beta})
  \gamma_\nu
  \Gamma 
 \Bigr]\; + 
 \nonumber\\[1mm]
 & + & \!\!\!
 \fr34 \gamma^\mu  (\bsl{P}_{\! 1} + m_{\nu_\alpha}) \gamma^\nu
 \sum_{\beta  = e,\mu,\tau}
 \tr\Bigl[ 
  (\bsl{P}_{\! 2} + m_{\nu_\beta})
  \gamma_\mu
  a_L 
  (\bsl{P}_{\! 3} + m_{\nu_\beta})
  \gamma_\nu
  a_L 
 \Bigr]\; - 
 \nonumber\\[1mm] 
 & - & \!\!\!
 \fr34 \gamma^\mu  (\bsl{P}_{\! 1} + m_{\nu_\alpha}) \gamma^\nu a_L 
  (\bsl{P}_{\! 2} + m_{\nu_\alpha})
  \gamma_\mu
  a_L 
  (\bsl{P}_{\! 3} + m_{\nu_\alpha})
  \gamma_\nu 
  \; - 
 \nonumber\\[2mm] 
 & - & \!\!\!
 \displaystyle
 \gamma^\mu  (\bsl{P}_{\! 1} + m_{l_\alpha}) \gamma^\nu
 \,\Gamma\,
  (\bsl{P}_{\! 2} + m_{l_\alpha})
  \gamma_\mu
  a_L 
  (\bsl{P}_{\! 3} + m_{\nu_\alpha})
  \gamma_\nu
  \; - 
 \nonumber\\[2mm] 
 & - & \!\!\!
 \gamma^\mu  (\bsl{P}_{\! 1} + m_{\nu_\alpha}) \gamma^\nu
 a_L
  (\bsl{P}_{\! 2} + m_{l_\alpha})
  \gamma_\mu
  \,\Gamma\, 
  (\bsl{P}_{\! 3} + m_{l_\alpha})
  \gamma_\nu 
  \biggr\} \times
  \nonumber \\[1mm] 
  & & \hspace*{0.5cm} \times
 (2\pi)^4 \delta^{(4)}(P_2-P_1-P_3-Q) \, 
 \nF{2}(1-\nF{1})(1-\nF{3}) + \ldots
 \;,  
 \hspace*{-0.2cm} \ScatF \hspace*{0.2cm}
 \la{pertL}
\ea
where now $\Gamma \equiv -1/2 + 2 x_\rmii{W} + \gamma_5/2$,  with
$x_\rmii{W} \equiv \sin^2 \theta_\rmii{W}$, and the  energies $E_i$
are on-shell with the obvious mass assignments. 

Inspecting \eqs\nr{master3} and \nr{pertL},  we observe that two
kinds of Dirac traces appear in the final result:  either a product
of two traces, or a single trace.  Both cases are elementary, and
result in
\ba
 \mathcal{T}_1 \!\! & \equiv & \!\! 
 \tr
 \Bigl[ 
   \bsl{E} a_L \gamma^\mu
   (\bsl{P}_{\! 1} + m_{1})
   \gamma^\nu a_R    
 \Bigr]
 \,
 \tr
 \Bigl[ 
   (\bsl{P}_{\! 2} + m_{2})
   \gamma_\mu 
   (a + b \gamma_5)
   (\bsl{P}_{\! 3} + m_{3})
   \gamma_\nu
   (a + b \gamma_5)
 \Bigr]
 \nonumber\\[1mm] 
 \!\! & = & \!\!
 16 \Bigl[ 
  (a-b)^2\; P_1 \cdot P_2 \; P_3 \cdot E + 
  (a+b)^2\; P_1 \cdot P_3 \; P_2 \cdot E + 
  (b^2 - a^2)\; m_2 m_3 \;P_1 \cdot E
 \Bigr]
 \;, \hspace*{0.5cm} \la{T1} \\[2mm]
 \mathcal{T}_2 \!\! & \equiv & \!\! 
 \tr
 \Bigl[ 
   \bsl{E} a_L \gamma^\mu
   (\bsl{P}_{\! 1} + m_{1})
   \gamma^\nu 
   (a + b \gamma_5)
   (\bsl{P}_{\! 2} + m_{2})
   \gamma_\mu 
   (c + d \gamma_5)
   (\bsl{P}_{\! 3} + m_{3})
   \gamma_\nu
   a_R 
 \Bigr]
 \nonumber\\[1mm] 
 \!\! & = & \!\! 
 8 \Bigl[ 
  -2 (a+b)(c+d)\; P_1 \cdot P_3 \; P_2 \cdot E + 
  (a+b)(c-d)\; m_2 m_3 \; P_1 \cdot E
 + \nonumber\\[1mm] & & 
 \hspace*{0.3cm} + 
  (a-b)(c-d)\; m_1 m_3 \; P_2 \cdot E
 + 
  (a-b)(c+d)\; m_1 m_2 \; P_3 \cdot E
 \Bigr]
 \;, \la{T2}
\ea
where the ``external'' four-vector $E$ is either $Q$ or $u$ 
(cf.\ \eq\nr{master3}).

Given these ingredients ---  the kinematic channels in
\eq\nr{pertB},  the Dirac structures in \eq\nr{pertL},  and the
traces in \eqs\nr{T1}, \nr{T2} ---  we can finally collect together
the full expression needed in \eq\nr{master3}. We obtain
\ba
 & &  \hspace*{-1cm}
 \tr\Bigl[\bsl{E} a_L \im\bsl{\Sigma}(Q) a_R \Bigr] = 
 2 G_F^2 \nF{}^{-1}(q^0)  \sum_i C_i  
 \int \! \frac{{\rm d}^3 \vec{p}_1}{(2\pi)^3 2 E_1} \,
 \int \! \frac{{\rm d}^3 \vec{p}_2}{(2\pi)^3 2 E_2} \,
 \int \! \frac{{\rm d}^3 \vec{p}_3}{(2\pi)^3 2 E_3} \,
 \times  
 \nn 
 & \times \biggl\{ & \!\!\!
 (2\pi)^4 \delta^{(4)}(P_1+P_2+P_3-Q )\, 
 \nF{1}\nF{2}\nF{3} \,
 \mathcal{T}_i(-m_1,m_2,-m_3)
 + \ScatA \nn & + & \!\!\!
 (2\pi)^4 \delta^{(4)}(P_2+P_3 -P_1-Q) \,
 \nF{2}\nF{3}(1-\nF{1}) \,
 \mathcal{T}_i(m_1,m_2,-m_3)
 + \ScatB \nn & + & \!\!\!
 (2\pi)^4 \delta^{(4)}(P_1+P_3-P_2-Q) \, 
 \nF{1}\nF{3}(1-\nF{2}) \,
 \mathcal{T}_i(-m_1,-m_2,-m_3)
 + \ScatC \nn & + & \!\!\!
 (2\pi)^4 \delta^{(4)}(P_1+P_2-P_3-Q) \, 
 \nF{1}\nF{2}(1- \nF{3}) \,
 \mathcal{T}_i(-m_1,m_2,m_3)
 + \ScatD \nn & + & \!\!\!
 (2\pi)^4 \delta^{(4)}(P_1-P_2-P_3-Q) \,
 \nF{1}(1-\nF{2})(1-\nF{3}) \,
 \mathcal{T}_i(-m_1,-m_2,m_3)
 + \ScatE \nn & + & \!\!\!
 (2\pi)^4 \delta^{(4)}(P_2-P_1-P_3-Q) \, 
 \nF{2}(1-\nF{1})(1-\nF{3}) \,
 \mathcal{T}_i(m_1,m_2,m_3)
 + \ScatF \nn & + & \!\!\!
 (2\pi)^4 \delta^{(4)}(P_3 -P_1-P_2-Q) \,
 \nF{3} (1-\nF{1})(1-\nF{2}) \,
 \mathcal{T}_i(m_1,-m_2,-m_3)
 + \ScatG \nn & + & \!\!\!
 (2\pi)^4 \delta^{(4)}(-P_1-P_2-P_3-Q ) \, 
 (1-\nF{1})(1-\nF{2})(1-\nF{3}) \,
 \mathcal{T}_i(m_1,-m_2,m_3)
 \biggr\}
 , \ScatH  \nn 
 \la{pertF}
\ea
where $\mathcal{T}_i$ equals $\mathcal{T}_1$ or $\mathcal{T}_2$, as 
specified in Table~1. The prefactors $C_i$ and the masses $m_1$,
$m_2$,  $m_3$ relevant for each channel are also listed in Table~1.
For completeness and future reference, we have included  in this
Table the perturbative hadronic contributions as well. 
 
\begin{table}[p]

\begin{minipage}[h]{15.5cm}
\begin{center}
\begin{tabular}{lll@{~~~}l@{~}l@{~}ll@{~~~}l@{~~~}l@{~~~}l}
\hline\hline
 channel & 
 $C_i$ & 
 $\mathcal{T}_i$ & 
 $m_1$&$m_2$&$m_3$ & 
 ~~~$a$&~~~$b$&~~~$c$&~~~$d$ \\
\hline\hline \\[-3mm]
 WW + hadrons & 
 $2 \Nc |V_{ud}|^2$ & 
 $\mathcal{T}_1$ & 
 $m_{l_\alpha}$&$m_d$&$m_u$ & 
 $+\fr12$&$-\fr12$&$ $&$ $ \\[2mm]
  & 
 $2 \Nc |V_{us}|^2$ & 
 $\mathcal{T}_1$ & 
 $m_{l_\alpha}$&$m_s$&$m_u$ & 
 $+\fr12$&$-\fr12$&$ $&$ $ \\[2mm]
  & 
 $2 \Nc |V_{cd}|^2$ & 
 $\mathcal{T}_1$ & 
 $m_{l_\alpha}$&$m_d$&$m_c$ & 
 $+\fr12$&$-\fr12$&$ $&$ $ \\[2mm]
  & 
 $2 \Nc |V_{cs}|^2$ & 
 $\mathcal{T}_1$ & 
 $m_{l_\alpha}$&$m_s$&$m_c$ & 
 $+\fr12$&$-\fr12$&$ $&$ $ \\[2mm]
 ZZ + hadrons & 
 ${\Nc}/{2}$ & 
 $\mathcal{T}_1$ & 
 $0$&$m_u$&$m_u$ & 
 $+\fr12 - \fr43 x_\rmii{W}$&$-\fr12$&$ $&$ $ \\[2mm]
  & 
 ${\Nc}/{2}$ & 
 $\mathcal{T}_1$ & 
 $0$&$m_c$&$m_c$ & 
 $+\fr12 - \fr43 x_\rmii{W}$&$-\fr12$&$ $&$ $ \\[2mm]
  & 
 ${\Nc}/{2}$ & 
 $\mathcal{T}_1$ & 
 $0$&$m_d$&$m_d$ & 
 $-\fr12 + \fr23 x_\rmii{W}$&$+\fr12$&$ $&$ $ \\[2mm]
  & 
 ${\Nc}/{2}$ & 
 $\mathcal{T}_1$ & 
 $0$&$m_s$&$m_s$ & 
 $-\fr12 + \fr23 x_\rmii{W}$&$+\fr12$&$ $&$ $ \\[2mm] \hline\\[-3mm]
 WW + leptons & 
 $2$ & 
 $\mathcal{T}_1$ & 
 $m_{l_\alpha}$&$m_e$&$0$ & 
 $+\fr12$&$-\fr12$&$ $&$ $ \\[2mm]
  & 
 $2$ & 
 $\mathcal{T}_1$ & 
 $m_{l_\alpha}$&$m_\mu$&$0$ & 
 $+\fr12$&$-\fr12$&$ $&$ $ \\[2mm]
  & 
 $2$ & 
 $\mathcal{T}_1$ & 
 $m_{l_\alpha}$&$m_\tau$&$0$ & 
 $+\fr12$&$-\fr12$&$ $&$ $ \\[2mm]
 ZZ + leptons & 
 ${1}/{2}$ & 
 $\mathcal{T}_1$ & 
 $0$&$m_e$&$m_e$ & 
 $-\fr12 + 2 x_\rmii{W}$&$+\fr12$&$ $&$ $ \\[2mm]
  & 
 ${1}/{2}$ & 
 $\mathcal{T}_1$ & 
 $0$&$m_\mu$&$m_\mu$ & 
 $-\fr12 + 2 x_\rmii{W}$&$+\fr12$&$ $&$ $ \\[2mm]
  & 
 ${1}/{2}$ & 
 $\mathcal{T}_1$ & 
 $0$&$m_\tau$&$m_\tau$ & 
 $-\fr12 + 2 x_\rmii{W}$&$+\fr12$&$ $&$ $ \\[2mm]
 ZZ + neutrinos & 
 ${9}/{4}$ & 
 $\mathcal{T}_1$ & 
 $0$&$0$&$0$ & 
 $+\fr12$&$-\fr12$&$ $&$ $ \\[2mm]
  & 
 $-{3}/{4}$ & 
 $\mathcal{T}_2$ & 
 $0$&$0$&$0$ & 
 $+\fr12$&$-\fr12$&$+\fr12$&$-\fr12$ \\[2mm]
 WZ + leptons & 
 $-1$ & 
 $\mathcal{T}_2$ & 
 $m_{l_\alpha}$&$m_{l_\alpha}$&$0$ & 
 $-\fr12+ 2 x_\rmii{W}$&$+\fr12$&$+\fr12$&$-\fr12$ \\[2mm]
  & 
 $-1$ & 
 $\mathcal{T}_2$ & 
 $0$&$m_{l_\alpha}$&$m_{l_\alpha}$ & 
 $+\fr12$&$-\fr12$&$-\fr12+ 2 x_\rmii{W}$&$+\fr12$ \\[2mm]
\hline\hline
\end{tabular}
\end{center}
\end{minipage}

\caption[a]{\small The coefficients and masses that  appear in
\eq\nr{pertF}, in the perturbative limit. The functions
$\mathcal{T}_1$ and $\mathcal{T}_2$, which depend on the masses
$m_1$, $m_2$, $m_3$ and the coefficients $a$, $b$, $c$, $d$ (which
take values as specified above),   are defined in \eqs\nr{T1} and
\nr{T2}, respectively. The symbols $l_\alpha$ stand for leptons of 
generation $\alpha$, i.e.\ 
$m_{l_1} \equiv m_e, m_{l_2} \equiv m_\mu, m_{l_3} \equiv m_\tau$.}

\end{table}

%
\section{Solution of the kinetic equation}
\la{se:kinetic}

Now that the ingredients entering \eq\nr{expansion} are in place, we
need to discuss the solution of this equation,  and identify the
precise information about the  equation-of-state which is needed for
the solution. The considerations that follow  are  rather standard
(see, e.g., refs.~\cite{Dodelson:1993je,Dolgov:2000ew}),  but we
present them here for completeness and in order to fix the notation.

It is convenient to represent the rate $R(T,q)$ as a function $F$ of
dimensionless variables,
\be
 R(T,q) = G_F^2 T^5 F\left(\frac{m_i}{T}, \frac{q}{T}\right) \;, 
\ee
where $m_i$ are the masses of the particles of the MSM and of the
sterile neutrinos. We also introduce the effective numbers of
massless bosonic degrees of freedom  $g_\rmi{eff}(T)$ and
$h_\rmi{eff}(T)$ via the relations
\be
 e(T) \equiv \frac{\pi^2 T^4}{30} g_\rmi{eff}(T) \;, \;\;\;
 s(T) \equiv \frac{2 \pi^2 T^3}{45} h_\rmi{eff}(T) \;,
\ee
where $e(T)$ and $s(T)$ are the energy and entropy densities, 
respectively. Given the equation-of-state of the plasma 
[i.e.\ the relation between the pressure and the temperature,
$p=p(T)$],  $g_\rmi{eff}(T)$ and $h_\rmi{eff}(T)$ can be found from 
the standard thermodynamical relations
\be
 g_\rmi{eff}(T)=\frac{30}{\pi^2 T^2}\frac{{\rm d}}{{\rm d}T}
 \left(\frac{p}{T}\right)
 \;, \quad
 h_\rmi{eff}(T)=\frac{45}{2 \pi^2 T^3} \frac{{\rm d}p}{{\rm d}T}
 \;.
 \la{heff}
\ee
Furthermore the sound speed squared,
$
 c_s^2(T)  = {p'(T)}/{e'(T)} = {p'(T)}/{Tp''(T)}
$, 
can be expressed as 
\be
 \frac{1}{c_s^2(T)} = 
 3 + \frac{T h'_\rmi{eff}(T)}{h_\rmi{eff}(T)}, 
\ee
where a prime denotes a derivative with respect to $T$.
The Hubble rate is given by
\be
 H=\frac{T^2}{M_0(T)}
 \;,\quad
 M_0(T)=M_\rmi{Pl} \left[\frac{45}{4\pi^3
 g_\rmi{eff}(T)}\right]^{\frac{1}{2}}\;,
\ee
where $M_\rmi{Pl}$ is the Planck mass.
Note that, from the assumption (iv) in \se\ref{se:intro}, 
the heavier sterile neutrinos do not contribute 
to $g_\rmi{eff}$ and $h_\rmi{eff}$ for the temperatures 
of interest. Furthermore, the contributions from the lightest 
sterile neutrino are negligibly small as long as it 
does not get thermalized.

The kinetic equation (\ref{kinetic}) can easily be integrated by
writing $n_I(t,q)=f(t,y)$, where the variable $y= a(t) q$ accounts
for red-shift. After simple manipulations one gets the distribution
function of sterile neutrinos at temperature $T_0$:
\be
 n_I(t_0,q)=\frac{G_F^2}{3} \int_{T_0}^\infty \! {\rm d}T \, 
 \frac{M_0(T)T^2}{c_s^2(T)}
 F\biggl( \frac{m_i}{T},
 \frac{q}{T_0}
 \left[\frac{h_\rmi{eff}(T)}{h_\rmi{eff}(T_0)}\right]^{\frac{1}{3}}
 \biggr)
 \;.
 \label{distribution}
\ee
The integral of \eq\nr{distribution} over the momenta $q$ gives the
number density of sterile neutrinos, which can  conveniently be
normalized with respect to its would-be equilibrium  value,
$n_\rmi{eq}(t_0) = 3 \zeta(3) T_0^3 / 2 \pi^2$:
\be
 \frac{n_I(t_0)}{n_\rmi{eq}(t_0)}=
  \frac{8 G_F^2 \pi^3}{9\zeta(3)}
  \int_{T_0}^\infty \! {\rm d}T
  \int_0^\infty \! {\rm d}z\, z^2 \, 
 \frac{M_0(T)T^2}{c_s^2(T)}
 \frac{h_\rmi{eff}(T_0)}{h_\rmi{eff}(T)}
 F\left(\frac{m_i}{T},z\right)
 \;.
 \label{concentration}
\ee
Finally another characterization of the same quantity  is obtained by
normalizing through the total entropy density,  which produces the
so-called yield parameter: 
\be
 Y_I(t_0) \equiv \frac{n_I(t_0)}{s(t_0)}
 = 
 \frac{30 G_F^2}{\pi}
 \int_{T_0}^{\infty}
 \! {\rm d} T 
 \int_0^{\infty}
 \! {\rm d}z \, z^2 \, \frac{M_0(T) T^2 }{c_s^2(T) h_\rmi{eff}(T)}
 F\left( \frac{m_i}{T},z\right)
 \;.
 \la{yield}
\ee 
The benefit of \eq\nr{yield} is that it obtains, unlike
\eq\nr{concentration}, a constant value at low temperatures (in the
absence of entropy production), because the factor $h_\rmi{eff}(T_0)$
drops out.

To summarise, to find the relic concentration of sterile neutrinos 
produced by active-sterile transitions, one should have a reasonable
approximation  for the hadronic equation-of-state $p(T)$ and for its 
first and second temperature derivatives, $p'(T)$ and $p''(T)$.

To convert the result of \eq\nr{yield} to physical units,  we denote
the contribution to the number density of  the lightest sterile
neutrino ($I=1$) from  an active neutrino of flavour $\alpha$, by
$n_{\alpha 1}$ (so that $n_1 = \sum_{\alpha = e,\mu,\tau} n_{\alpha
1}$).  To relate this quantity to $\Omega_\rmi{dm}$, we write 
\be
 \Omega_{\alpha 1} \equiv \frac{M_1 n_{\alpha 1}}{\rho_\rmi{cr}} = 
          \frac{M_1 Y_{\alpha 1}}{\rho_\rmi{cr}/s}
 \;,
\ee
where 
$
 Y_{\alpha 1} \equiv {n_{\alpha 1}}/{s}
$.
{}From Particle Data Group~\cite{pdg}, one finds 
$
 \rho_\rmi{cr} \approx 1.054 \times 10^{-5}\, h^2 \, \mbox{GeV}\,\mbox{cm}^{-3}
$
and 
$
 s = 7.04\, n_\gamma \approx 2886\, \mbox{cm}^{-3}
 \;,
$
yielding
$
 {\rho_\rmi{cr}}/{s} \approx 3.65 \times 10^{-9} h^2 \, \mbox{GeV}
 \;.
$
Then the result of \eq\nr{yield} can be expressed as
\be
 \Omega_{\alpha 1} h^2 = 0.11\, C_\alpha(M_1) \, 
 \biggl( \frac{|M_D|_{\alpha 1}}{0.1\;\mbox{eV}} \biggr)^2
 \;, \la{Ca}
\ee
where
$
 C_\alpha (M_1) = 2.49\times 10^{-5} \times Y_{\alpha 1}/\theta_{\alpha 1}^2 
 \times (\mbox{keV} / M_1)
$.

The total relic density of the lightest sterile neutrino is 
now given by $\sum_{\alpha = e,\mu,\tau}  \Omega_{\alpha 1} h^2$.
On the other hand, the dark-matter density in the present Universe
has been presicely measured by 
the WMAP collaboration~\cite{Spergel:2006hy} (68\% CL):
\begin{eqnarray}
  \Omega_{\rm dm} h^2 = 0.105^{+0.007}_{-0.013} \,.
\end{eqnarray}
Therefore, to avoid the overclosure of the Universe by $N_1$,
we get 
$
 \sum_{\alpha = e,\mu,\tau}  \Omega_{\alpha 1} h^2 \lsim 0.112
$,
which leads to
\be
 \sum_{\alpha = e,\mu,\tau} C_\alpha(M_1) \, 
 \biggl( \frac{|M_D|_{\alpha 1}}{0.1\;\mbox{eV}} \biggr)^2
 \lsim 1.0
 \;. \la{constr0}
\ee
This corresponds to \eq\nr{main}. In the following, we will present
our results for the functions $C_\alpha(M_1)$ defined by \eq\nr{Ca}.

%
\section{Hadronic uncertainties}
\la{se:qcd}

As discussed in ref.~\cite{als}, there are in principle three 
sources of hadronic effects and corresponding hadronic uncertainties
in the present computation:  those entering $\re\Sigma$, $\im\Sigma$,
as well  as the overall equation-of-state through the considerations
in \se\ref{se:kinetic}.  The hadronic contributions to $\re\Sigma$
are suppressed by $\alpha_w$ with respect to the leptonic ones and
will be omitted in the following.  The hadronic contributions to
$\im\Sigma$ arise at the same order as the leptonic ones, consituting
thus a significant  source term. Even more important are the
hadronic  contributions to the equation-of-state, since in the 
temperature range of interest hadrons completely  dominate over
leptons as far as the change in the effective  number of light
degrees of freedom is concerned  (cf.~\fig\ref{fig:eweos}).  On the
other hand, the hadronic effects in the equation-of-state  are
understood somewhat better than the ones in $\im\Sigma$, given  that
they are theoretically more straightforward to access, both in the
framework of the weak-coupling expansion~\cite{gsixg}--\cite{pheneos}
and of lattice simulations~\cite{Nfcp}--\cite{Nfbi},  so that the
{\em relative}  uncertainties are perhaps smaller then in
$\im\Sigma$.  In fact we estimate the {\em absolute} uncertainties
from these  two sources to be of similar magnitudes. In the
following  we discuss the ways in which we extract these two
quantities and try to estimate their uncertainties. 

\begin{figure}[t]

\centerline{%
\epsfysize=7.5cm\epsfbox{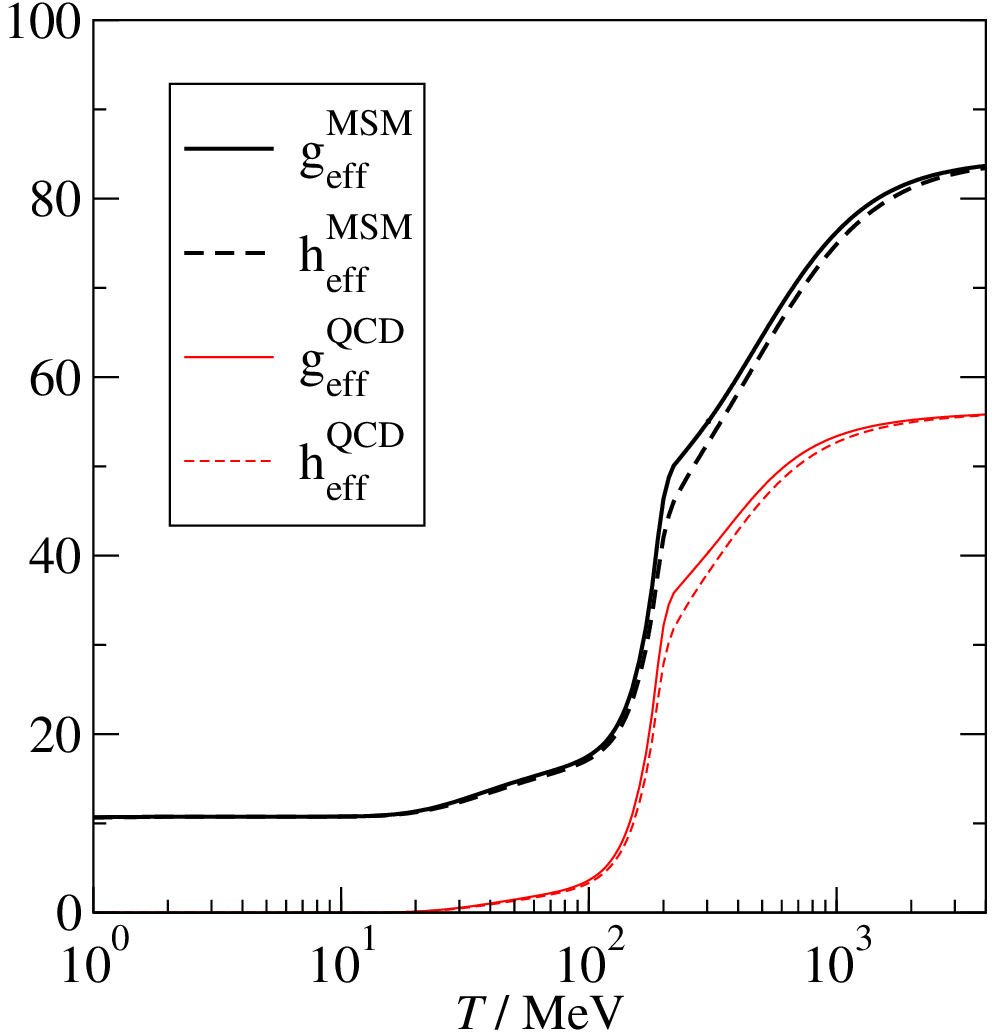}%
~~~~~~\epsfysize=7.5cm\epsfbox{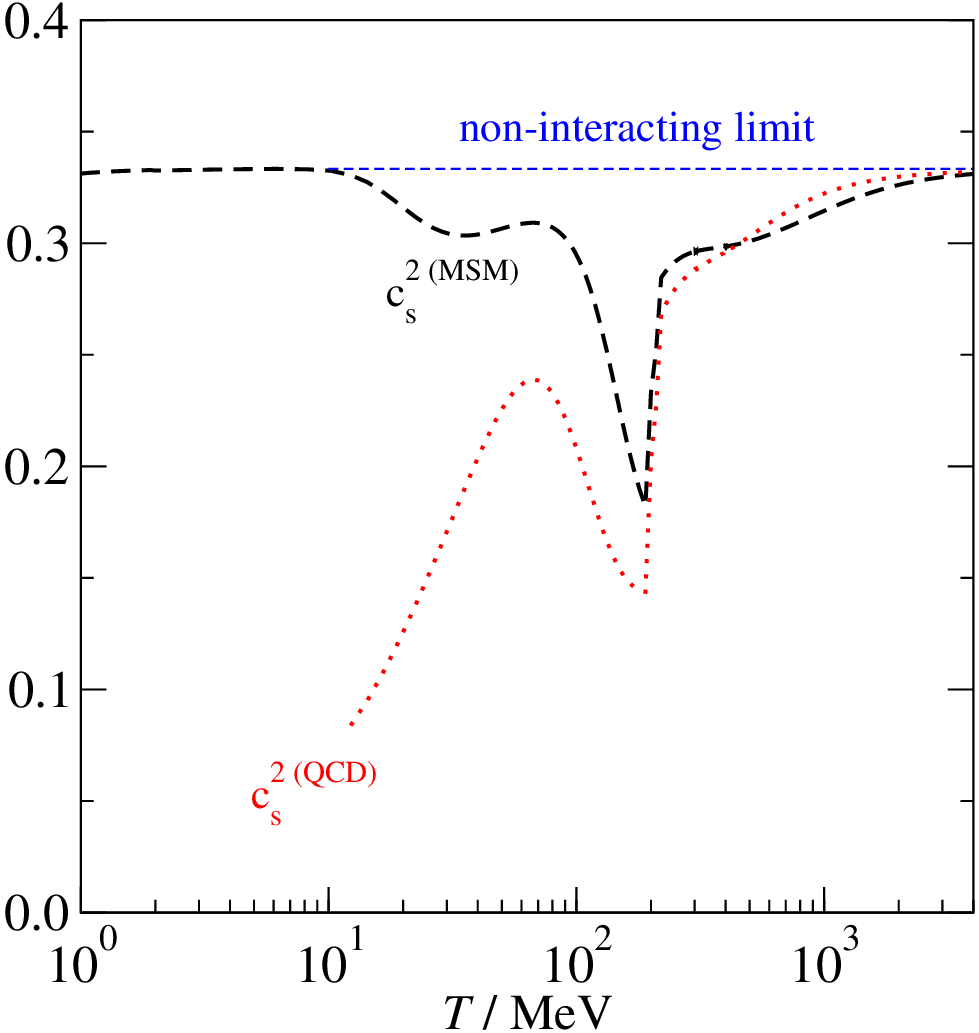}%
}

\caption[a]{\small
Left:  $g_\rmi{eff}, h_\rmi{eff}$ as defined 
in~\eq\nr{heff}, for the MSM and for the QCD part thereof, 
for $T = 1$~MeV \ldots 4~GeV
(for more details, see ref.~\cite{pheneos}).
Right: the speed of sound squared $c_s^2$, for the same systems.
Various sources of uncertainties in these estimates
are discussed in the text.
} 

\la{fig:eweos}
\end{figure}

\subsection{QCD equation-of-state}
\la{se:qcdeos}

A basic fact to realise about the QCD equation-of-state  is that the
low-temperature hadronic world and the high-temperature partonic
world can be analytically connected to each other,  by tuning quark
masses and the temperature:  at non-zero quark masses there is no
order parameter to  distinguish between the two situations. Moreover,
lattice simulations have been suggesting for quite a while already
that such an analytical  ``crossover'' is met even if the quark
masses are {\em not} tuned but kept  at their physical values; for a
recent studies see, e.g., refs.~\cite{fp}. In fact, these statements
could even have been made without recourse to  lattice simulations:
the thermodynamics of a dilute ``gas'' of  hadronic resonances at low
temperatures, and of {\em interacting}   quarks and gluons at high
temperatures, appear to extrapolate  to each other surprisingly
well,  when determined precisely enough~\cite{pheneos}.

Now, even if there only were a smooth crossover in the system,  it is
conventional to assign a (pseudo)critical temperature to it;  we will
denote this temperature by $\Tc$.  Of course $\Tc$ cannot be defined
precisely, but the ``conceptual'' ambiguities involved do not appear 
to be much larger than the current statistical uncertainties. 
A recent large-scale lattice study, for instance, suggests 
that $\Tc \simeq 192 \pm 8$~MeV~\cite{Cheng:2006qk}. 

On the other hand, it is also important to realise that at the 
current moment the lattice studies still involve {\em systematic}
uncertainties (related for instance to finite-volume effects or 
the absence of a precise continuum extrapolation), 
which are not quantitatively under control. Consequently, 
even though the final word on the value of $\Tc$ and  on what happens
around it lies with the lattice simulations, they are 
still rather far from
establishing the correct behaviour in a wide temperature interval. 
In fact, even the results  of various groups employing similar
techniques differ by much more than the statistical uncertainties cited
above~\cite{Aoki:2006br}. Moreover, lattice studies  fail to show any
signs of the characteristic peak that physical pions  cause in the
sound speed at $T \approx 70$~MeV~\cite{pheneos} (see also
\fig\ref{fig:eweos}),  a problem which can probably be assigned to
the fact that the quark  discretizations used in the current
finite-temperature simulations  do not respect the chiral symmetry
that plays an essential role at low temperatures, and also tend to
employ unphysically heavy quark masses. Lattice simulations also
cannot be applied at very high temperatures. To summarise, the
current numbers can be  expected to be fairly reliable in the range
from  about $\Tc$ to about $(2...3)\Tc$ at best.

For these reasons, we prefer to  adhere to the procedure introduced
in ref.~\cite{pheneos} here,  rather than to lattice simulations. It
makes use of a gas of hadronic resonances at low temperatures;  the
most advanced (up to resummed 4-loop level~\cite{gsixg})
weak-coupling  results at high temperatures; and an interpolation
thereof at intermediate temperatures.\footnote{%
  We have corrected a minor error in  the numerical results of
  ref.~\cite{pheneos}.} 
Remarkably, the temperature interval where an interpolating function
is needed in order to sew together the two asymptotic functions is
fairly narrow, not more than $10-20$~MeV, and centered around 
$T \approx 200$~MeV. We will refer to this temperature in the 
following as $\tilde\Tc$, but stress that, 
despite the curiously good agreement with the crossover
temperature suggested by the lattice study mentioned~\cite{Cheng:2006qk},  
$\tilde\Tc$ does not need to coincide with any specific definition of $\Tc$, 
given the inherent ambiguity in the location of a crossover. 
The significant benefit of this
recipe, compared with lattice  simulations, is that the results can
be evaluated also at arbitrarily low and high temperatures without
problems. 

Now, this recipe is naturally not exact in the intermediate
temperature range; the results can be improved when future lattice 
simulations get closer to the infinite-volume, 
continuum, and chiral limits. In order
to estimate the uncertainties in the present implementation, we have
considered two types of variations of the basic setup:  
\begin{itemize}

\item 
The shape and ``sharpness'' of the crossover can be changed by  adding
artificially heavy hadronic resonances to the gas,  or removing
existing ones.

\item
The overall location of the pseudocritical temperature can  be
changed most simply by just rescaling the temperature units by some
percentual amount. 

\end{itemize}
It is naturally not difficult to come up with other possible
variations as well, but as such recipes are purely phenomenological
in any case, it is sufficient for our purposes to restrict to these
most straightforward possibilities. 

We find that the first of these variations has a relatively minor
impact on the results. On the contrary, a change in $\tilde\Tc$ would lead
to visible changes. Qualitatively, a low $\tilde\Tc$ implies a low 
abundance, since the results are inversely proportional  to
$g_\rmi{eff}^{1/2} h_\rmi{eff}$ (cf.~\eq\nr{yield}), and the
production then peaks  on the partonic side where $g_\rmi{eff}^{1/2}
h_\rmi{eff}$ is higher.  Consequently, we will assume in the
following that  the uncertainties of the equation-of-state can be
sufficiently  estimated by a rescaling of $\tilde\Tc$ by 20\% in both
directions, $\tilde\Tc$ $= (160...240)$~MeV,  
which should be a conservative estimate.  

\subsection{Vector and axial current spectral functions}

As has been discussed in ref.~\cite{als}, the hadronic  contributions
to $\im\Sigma$ can be expressed through the spectral functions
related to  vector and axial currents with various flavour
structures.  The current status of the in general very  demanding
challenge of reliably determining these  quantities was also reviewed
in ref.~\cite{als}.

Irrespective of all details, however,  the general pattern we expect
for the hadronic effects should be clear:   at high temperatures $T
\gg \Tc$, hadrons should be reasonably represented by free quarks
(cf.\ \eq\nr{pertB}), whereas once the temperature is lowered,
confinement sets in, and all hadronic effects  eventually switch off,
given that all mesons and  baryons are massive. For $T \ll \Tc$, for
instance,  the hadronic spectral functions can be determined with
chiral perturbation  theory. It is easy to see that  to leading order
in the chiral expansion and assuming a mass-degenerate limit, only
the flavour non-singlet axial  current, denoted by $A^a_\mu$ in
ref.~\cite{als}, contributes.  We have verified explicitly that  this
dominant pionic contribution is strongly suppressed with respect to
the  leptonic contributions at all relevant 
temperatures (below 1\%),  for the
small masses $M_1$ that we are interested in.  

To give another
argument in the same direction,  we may reasonably assume that the
hadronic effects  are proportional to the number of effective
hadronic degrees of freedom,   $h_\rmi{eff}^\rmi{QCD}$,  related to
the hadronic contribution to the entropy, which  also rapidly
decreases with the temperature~(cf.\ \fig\ref{fig:eweos}).

These considerations suggest that a strict {\em upper bound}  for the
hadronic contribution to $\im\Sigma$ can be obtained by simply
computing the effect of free quarks, \eq\nr{pertB}. We take into
account the $u,d,s$ and $c$ quarks, cf.\ Table~1,  with their
$\msbar$ scheme masses~\cite{pdg}.
A strict {\em lower bound} can obviously be obtained
by simply setting $\Nc = 0$ in the hadronic contributions listed in
Table~1.  As we have already mentioned, the ``error bars'' that result
from this (clearly most conservative) procedure are roughly  of the
same order as  those related to the equation-of-state,  as defined in
\se\ref{se:qcdeos}.  (In fact they are a bit larger as we will see, 
but the recipe is also overly conservative.) Furthermore we define a
phenomenological  ``mean'' value by making use of the perturbative
contribution in Table~1,  but rescaling $\Nc$ by the factor
$h_\rmi{eff}^\rmi{QCD}(T)/58$,  where 58 counts the hadronic degrees
of freedom in  the non-interacting four-flavour limit.
%

%
\section{Numerical results for the sterile neutrino abundance}
\la{se:numerics}

For the numerical evaluation of the sterile neutrino abundance it is
useful to note that the computationally most  demanding part entering
our equations, $\im\Sigma$, is practically independent of $M_1$ in
the range $M_1 \ll T$ that we are  interested in. Therefore,
$\im\Sigma$ can be evaluated once and for all by using any fixed
$M_1$.  In contrast, the ``prefactor'' (i.e.\ the parts of
\eq\nr{master3} other than $\im\Sigma$) does depend strongly on the
precise value of $M_1$, yielding a non-trivial dependence on $M_1$
for the function $C_\alpha(M_1)$ of \eq\nr{Ca}.

As another comment, we note that for $q^0 > 0$ there are  seven
channels in \eq\nr{pertF} that can be realised. It can be verified
numerically, however, that in practice only the $2\to 2$ processes
give  a significant contribution; the other cases have a restricted
phase space and amount to at most 5\% of the $2\to 2$ channels
(for typical parameter values in fact much less than 5\%), which
is smaller than our uncertainties. To decrease the numerical cost, we
have therefore omitted these contributions in the following. It may
be noted, though,  that these additional channels systematically {\em
increase} the  sterile neutrino abundance.  

The way we have numerically evaluated the phase space integrations
entering $\im\Sigma$ at the 2-loop level is  explained in some detail in
Appendix~B. Once an efficient implementation is  available, the
remaining integrations (over $T,q$; cf.\ \eq\nr{yield}) are 
relatively straightforward. We note that the $q$-integration
converges exponentially fast, and can in practice be limited to 
values $q/T \lsim 10 ... 15$.

\begin{figure}[t]


\centerline{%
\epsfysize=9.0cm\epsfbox{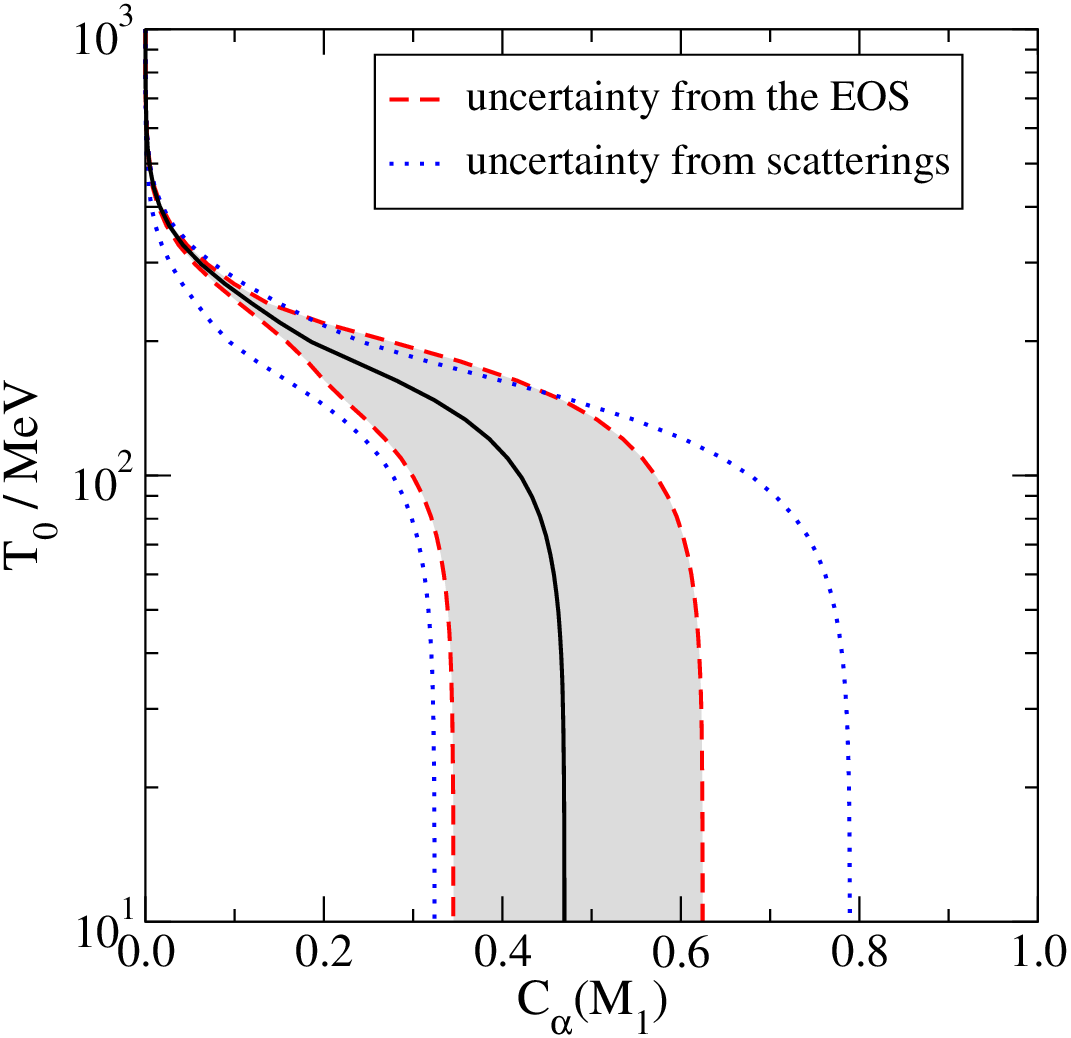}%
}

\caption[a]{\small An example of the 
$T_0$-evolution of $C_\alpha$  (cf.\ \eqs\nr{yield}--\nr{Ca}), 
for a fixed $\alpha = e$ and $M_1 = 10$~keV. 
Shown are the two sources of hadronic uncertainties
that are discussed in the text: 
from the equation-of-state (EOS) and from hadronic scatterings.} 
\la{fig:Tevol}
\end{figure}

In \fig\ref{fig:Tevol} we show an example of the $T_0$-evolution 
of the function $C_\alpha(M_1)$, as defined by \eq\nr{Ca}. We observe
that most of the sterile neutrino abundance is indeed generated 
at temperatures of a few hundred MeV for $M_1$ in the keV range.

\begin{figure}[t]


\centerline{%
\epsfysize=7.5cm\epsfbox{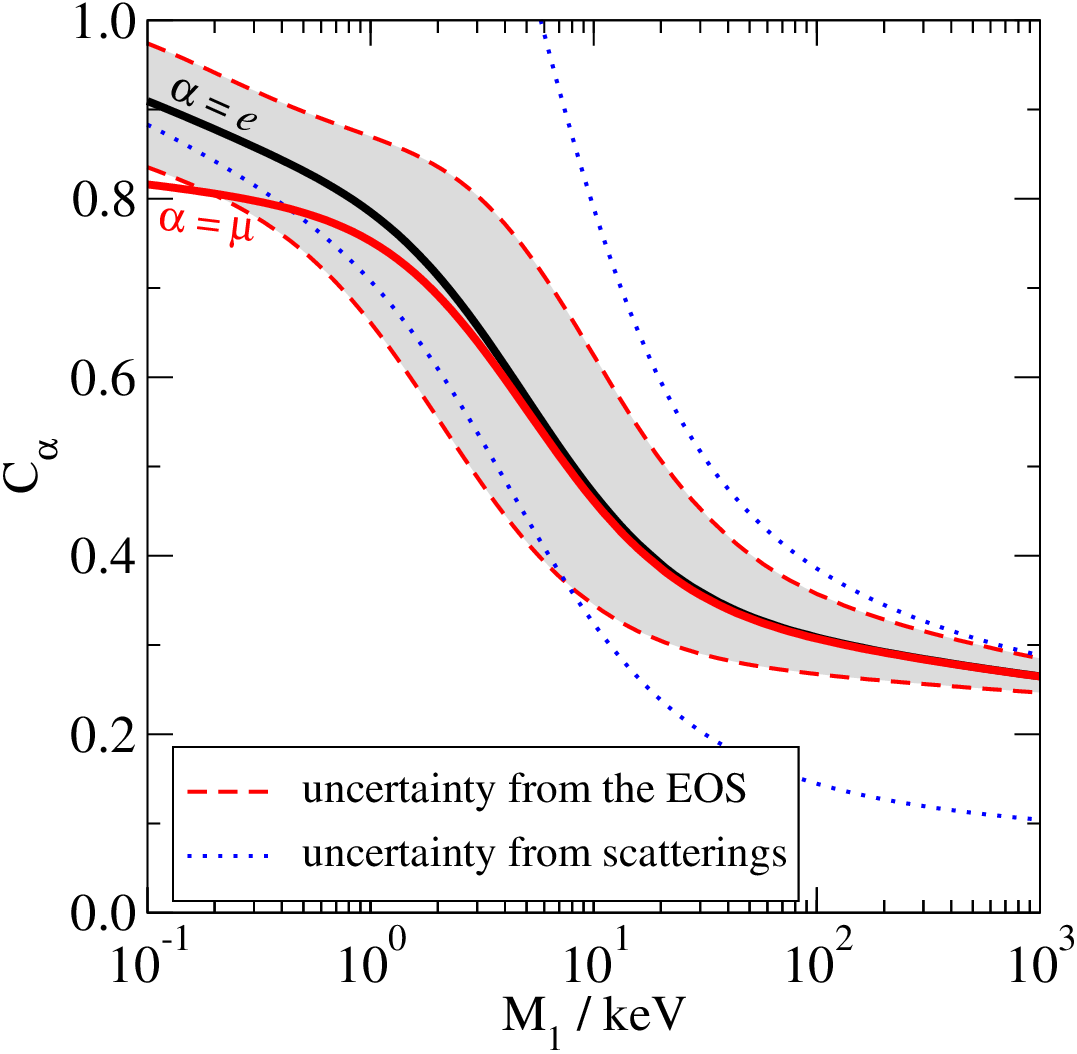}%
~~~~\epsfysize=7.5cm\epsfbox{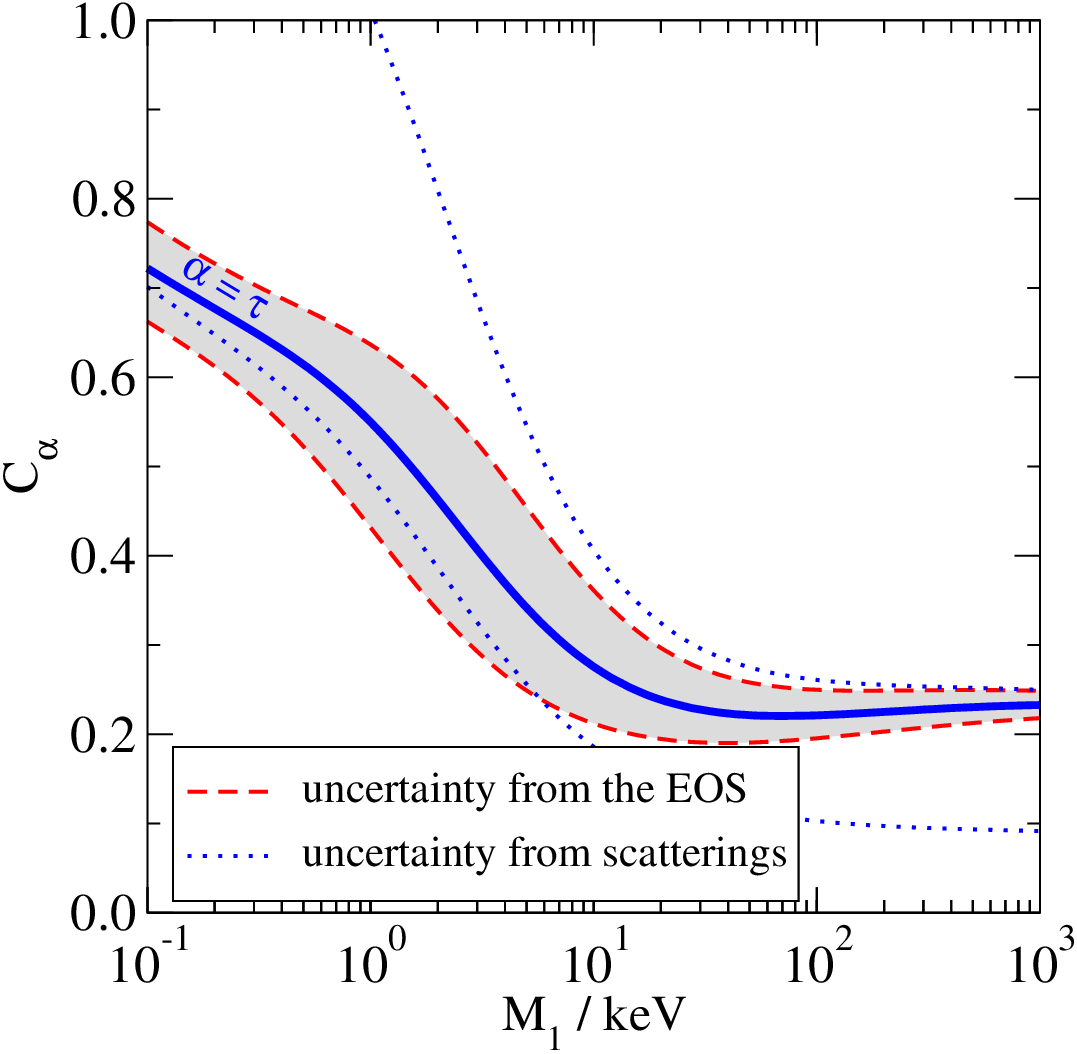}%
}

\caption[a]{\small The functions $C_\alpha(M_1)$, together with 
estimated hadronic uncertainties, as described in 
the text (see also \fig\ref{fig:Tevol}). 
Left: $\alpha = e,\mu$, and only the uncertainties for 
$\alpha =e$ are shown.
Right: $\alpha = \tau$.} 
\la{fig:Calpha}
\end{figure}

In \fig\ref{fig:Calpha} the final low-temperature values of 
$C_\alpha(M_1)$ are shown as a function of $\alpha$ and $M_1$. 
Importantly, we observe that the dependence on $M_1$ is fairly mild 
for a wide range of $M_1$, say $M_1=(0.1$--$10^3)$ keV,
once expressed as in \eq\nr{Ca}. This means, roughly speaking,  
that the relic density of $N_1$ is determined by the Dirac neutrino 
masses $|M_D|_{\alpha 1}$. According to \eq(\ref{constr0}) the Dirac neutrino 
masses should typically be $|M_D|_{\alpha 1} \lsim  0.1$ eV.

Furthermore, due to the differences between $C_\alpha$, 
the relic abundance of $N_1$ depends on the flavour structure 
of the Dirac neutrino masses $|M_D|_{\alpha 1}$.
We find the (moderate) hierarchy $C_e > C_\mu > C_\tau$.
Thus, the largest abundance is obtained when
\begin{eqnarray}
  | M_D |_{e 1} \neq 0\,,~~
  | M_D |_{\mu 1} = | M_D |_{\tau 1} = 0 \,.
  \la{case1}
\end{eqnarray}
In this case (which we call ``case 1''), we obtain the most stringent
upper bound on $\theta^2$ from \eq(\ref{constr0}).  On the other hand, 
the case (which we call ``case 2'') when 
\begin{eqnarray}
  | M_D |_{\tau 1} \neq 0\,,~~
  | M_D |_{e 1} = | M_D |_{\mu 1} = 0 \,,
  \la{case2}
\end{eqnarray}
gives the smallest relic abundance and the weakest upper bound on 
$\theta^2$. 

\begin{figure}[t]


\centerline{%
\epsfysize=9.0cm\epsfbox{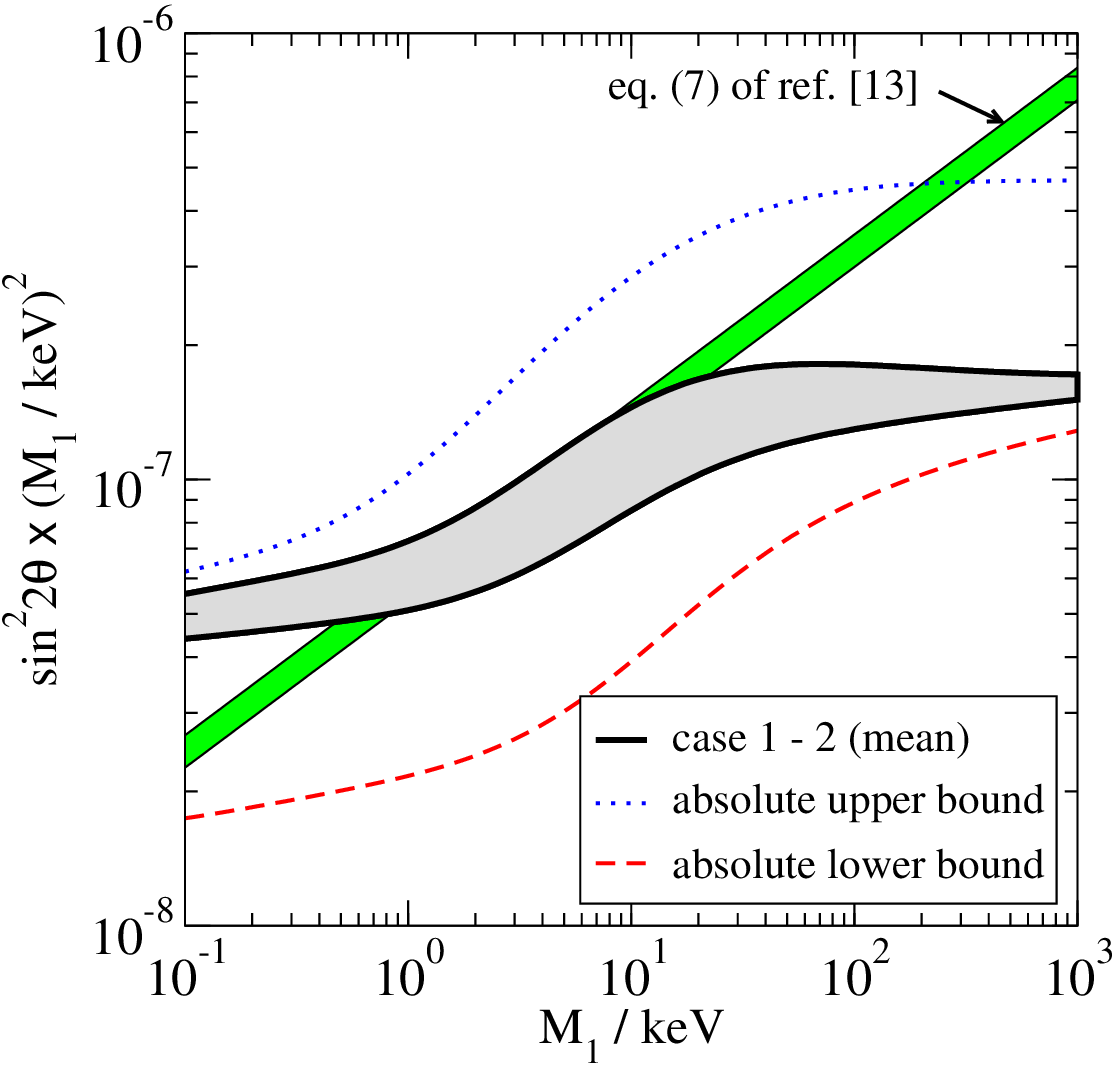}%
}

\caption[a]{\small 
The parameter values that, according to our
theoretical computation (cf.\ \fig\ref{fig:Calpha} and \eq\nr{Ca}), 
lead to the correct dark matter abundance in the Dodelson-Widrow scenario;
if additional sources are present, $\sin^2\!2\theta$ must lie 
{\em below} the curves shown.  The grey region between
case 1 (lower solid line) and 
case 2 (upper solid line) corresponds to different 
patterns of the active-sterile mixing angles, cf.\ \eqs\nr{case1}, \nr{case2}.
The absolute upper and 
lower bounds correspond to one of these limiting
patterns with simultaneously 
the uncertainties from the EOS and from hadronic scatterings 
set to their maximal values. The yellow band indicates 
the result in \eq(7) 
of ref.~\cite{Abazajian:2005gj},\footnotemark[2]
where we have inserted 
$\Omega_\rmi{dm} = 0.20$, and varied the parameter $\Tc$
in the range $\Tc = (150 ... 200)$~MeV. 
} 
\la{fig:exclusion_th}
\end{figure}

In Fig.~\ref{fig:exclusion_th}, we show the upper bounds on 
the mixing angle for the above two cases. We also compare with
the most recent previous computation from 
the literature~\cite{Abazajian:2005gj}.
It can be seen that the bound obtained
is rather insensitive to the flavour 
structure of the Dirac neutrino masses $|M_D|_{\alpha 1}$, 
at least when plotted on a logarithmic scale.

If we ignore the dependence of  $C_\alpha$ 
on $M_1$ and $\alpha$ in \fig\ref{fig:Calpha},
setting $C_\alpha \simeq 0.5$, we obtain a very simple but 
useful approximate bound, 
\be
 \sin^2(2\theta)
 \approx 
 4 \! \sum_{\alpha = e,\mu,\tau}\theta_{\alpha 1}^2 
 \; \lsim \; 
 8 \times 10^{-8} 
 \,\biggl( \frac{M_{1}}{\mbox{keV}} \biggr)^{-2}
 \;. \la{constr1}
\ee
More precise expressions, based on fitting 
the numerical data, will be given in \eqs\nr{fits} of the next section.
In~\fig\ref{fig:exclusion_th} the units on the $y$-axis 
have been so chosen that a direct comparison with the
approximate formula in \eq\nr{constr1} is possible.

\begin{figure}[t]


\centerline{%
\epsfysize=7.5cm\epsfbox{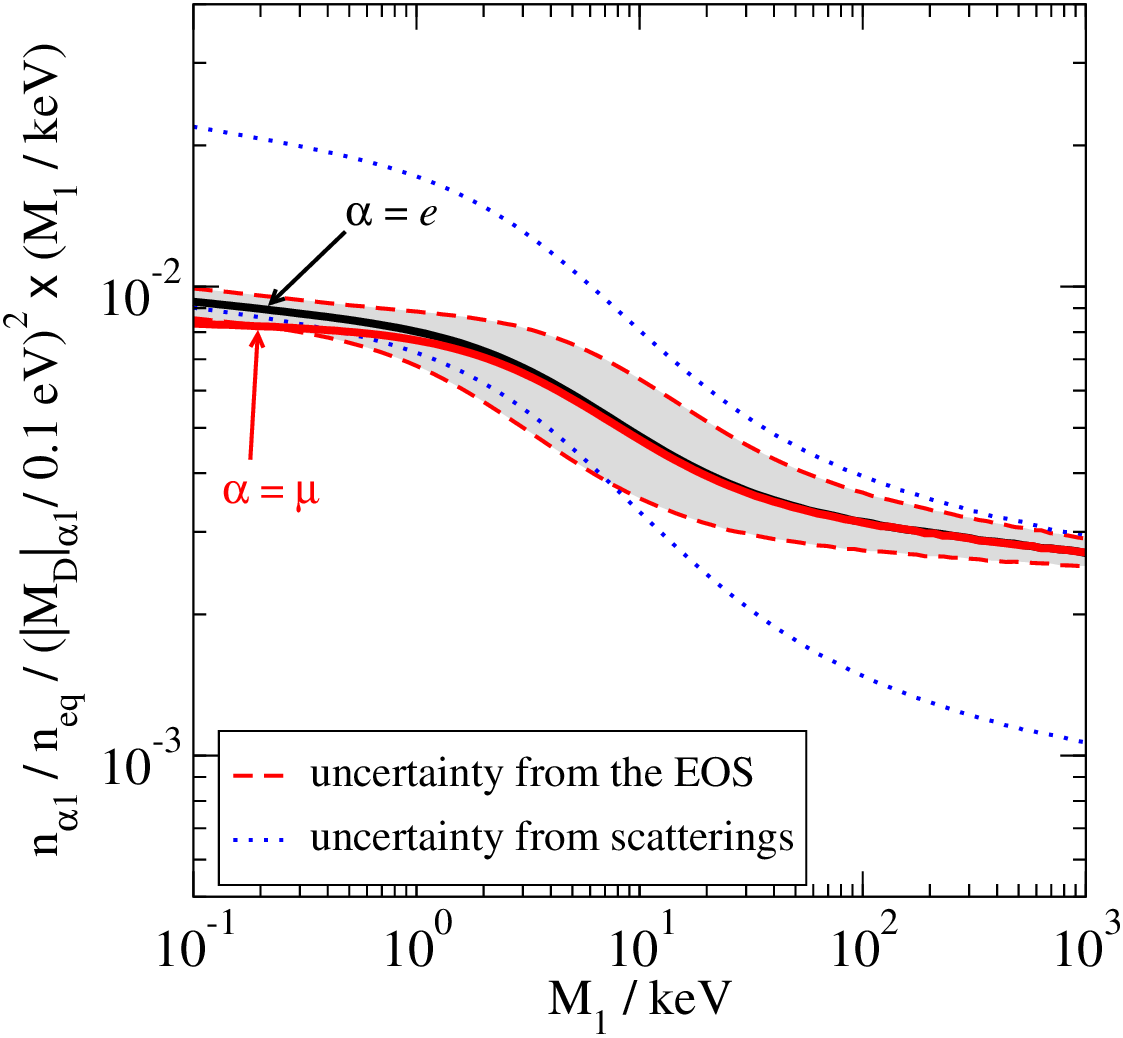}%
~~~\epsfysize=7.5cm\epsfbox{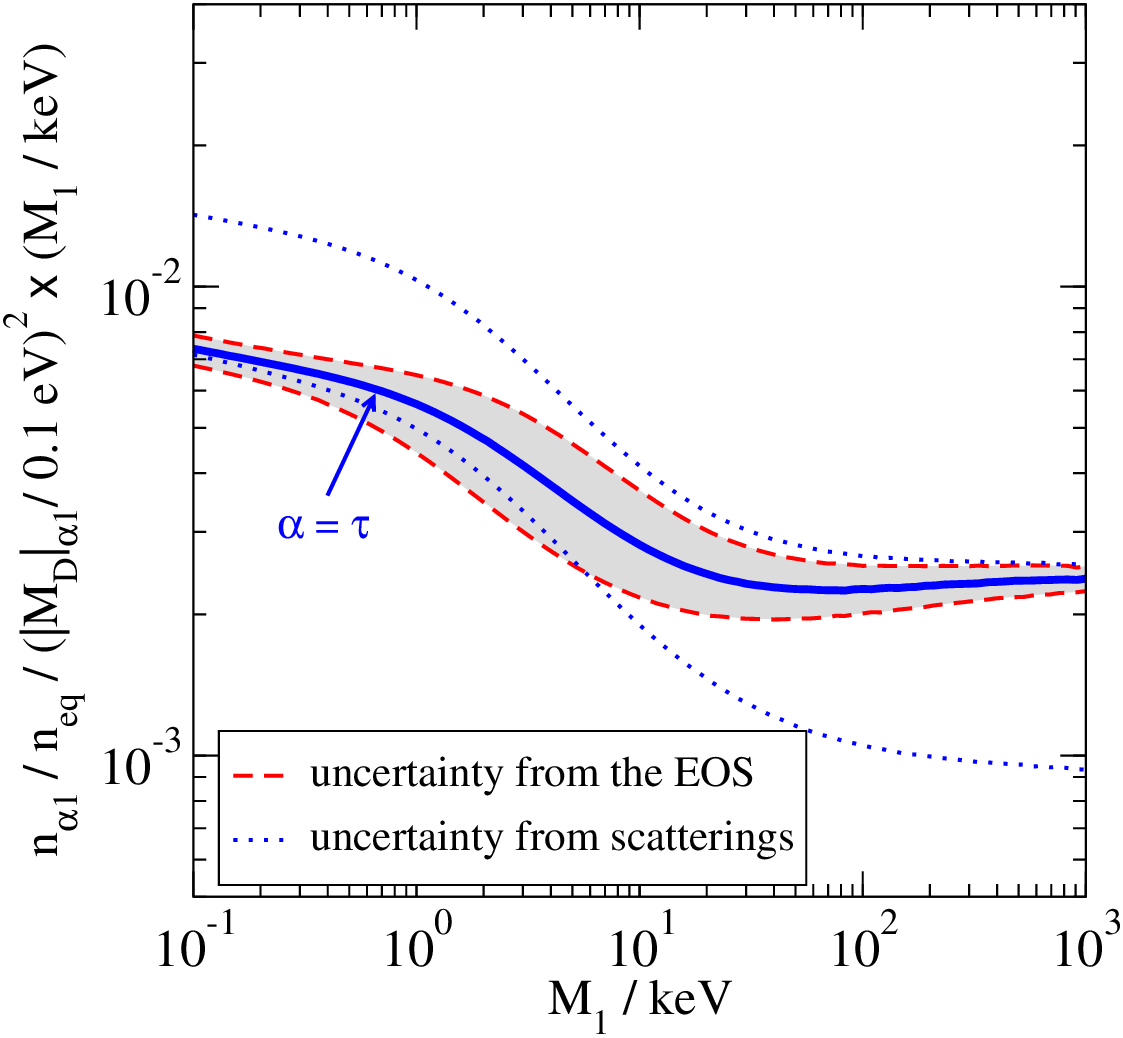}%
}

\caption[a]{\small The sterile neutrino abundance normalised to its 
equilibrium abundance, $n_{\alpha 1}/n_\rmi{eq}$, at $T_0 = 1$~MeV. 
Hadronic uncertainties have the same meaning as in  \fig\ref{fig:Tevol}.
The combination $( {|M_D|_{\alpha 1}} / {0.1\;\mbox{eV}} )^2$ 
has been factored out in analogy with \eq\nr{Ca} and, for better visibility, 
the results have been multiplied by $M_1 / $keV.
Left: $\alpha = e,\mu$, and only the uncertainties for $\alpha = e$ 
are shown.
Right: $\alpha = \tau$.} 
\la{fig:noneq}
\end{figure}

An alternative normalization of the sterile neutrino abundance  is
given in \eq\nr{concentration}, and the corresponding results are
plotted in \fig\ref{fig:noneq}. It is observed that the  abundance
generated is typically  much below its equilibrium value, as expected
[recall that in order to avoid overclosure, 
$
 ( {|M_D|_{\alpha 1}} / {0.1\;\mbox{eV}} )^2 \lsim 
$ a few, cf.\ \eq\nr{constr0} and \fig\ref{fig:Calpha}].

\begin{figure}[t]


\centerline{%
\epsfysize=7.5cm\epsfbox{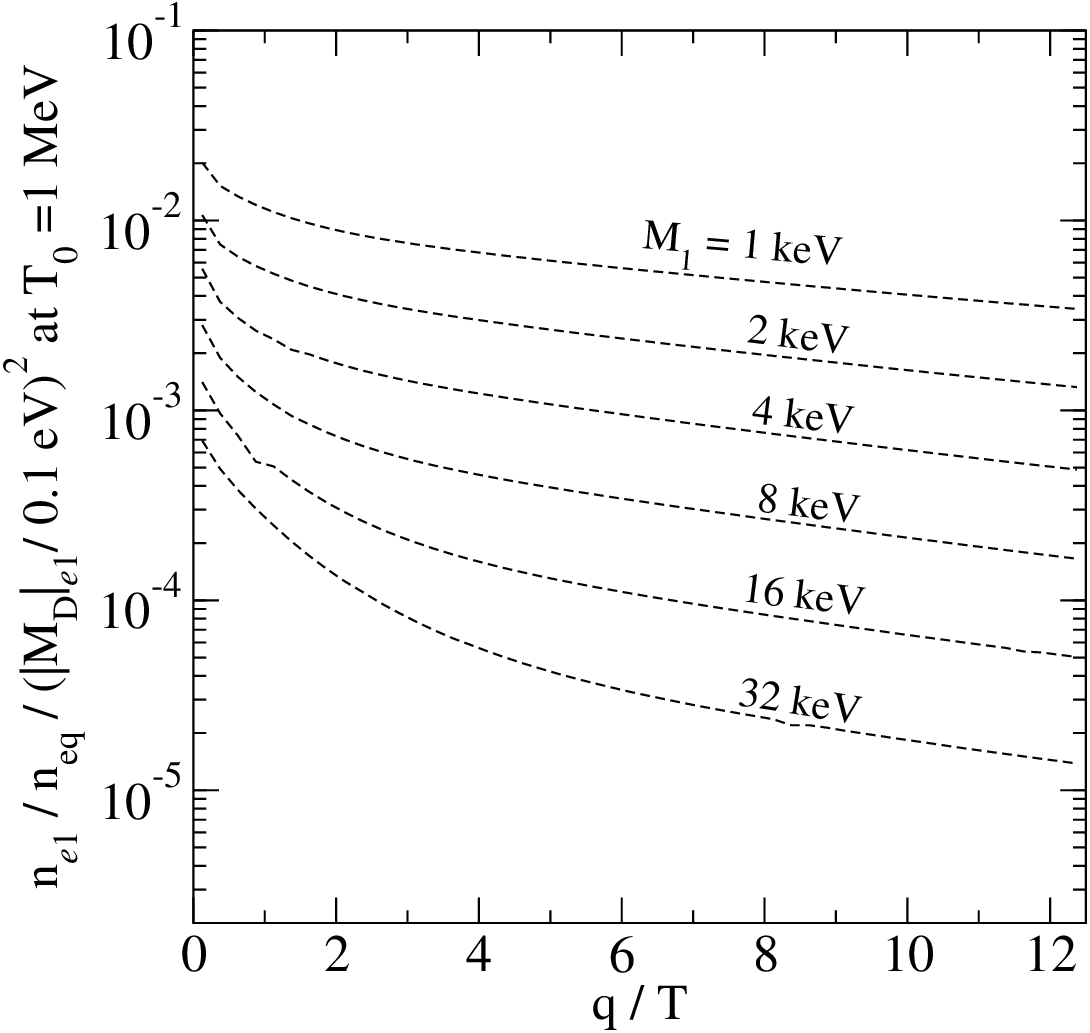}%
~~~\epsfysize=7.5cm\epsfbox{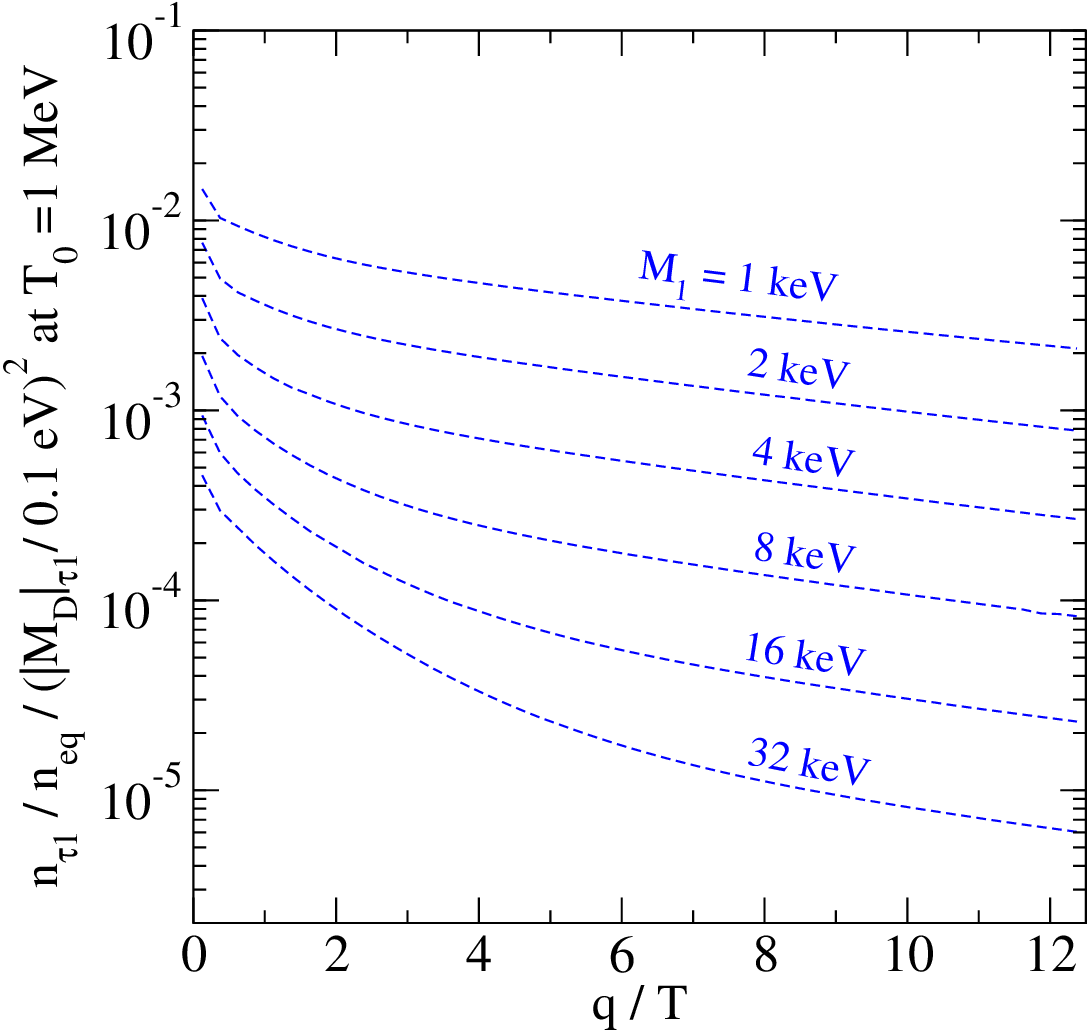}%
}

\caption[a]{\small  
The distribution functions $n_{\alpha 1}(t_0,q)$,
for $T_0 = 1$~MeV,  normalised to the massless equilibrium 
case, $n_\rmi{eq}(t_0,q) = 2 \nF{}(q) /(2\pi)^3$.
The combination $( {|M_D|_{\alpha 1}} / {0.1\;\mbox{eV}} )^2$ 
has been factored out in analogy with \eq\nr{Ca}.
Left: $\alpha = e$.
Right: $\alpha = \tau$.
These results can be compared with \fig{1} of ref.~\cite{Abazajian:2005gj}, 
showing the same distribution functions for 0.3~keV $\le M_1 \le$ 140~keV; 
although similar at first sight, there are significant differences on closer
inspection, for instance our curves are monotonic functions of $q/T$ 
even for small $M_1$, unlike those in  ref.~\cite{Abazajian:2005gj}.
} 
\la{fig:distr}
\end{figure}

\begin{figure}[t]


\centerline{%
\epsfysize=7.5cm\epsfbox{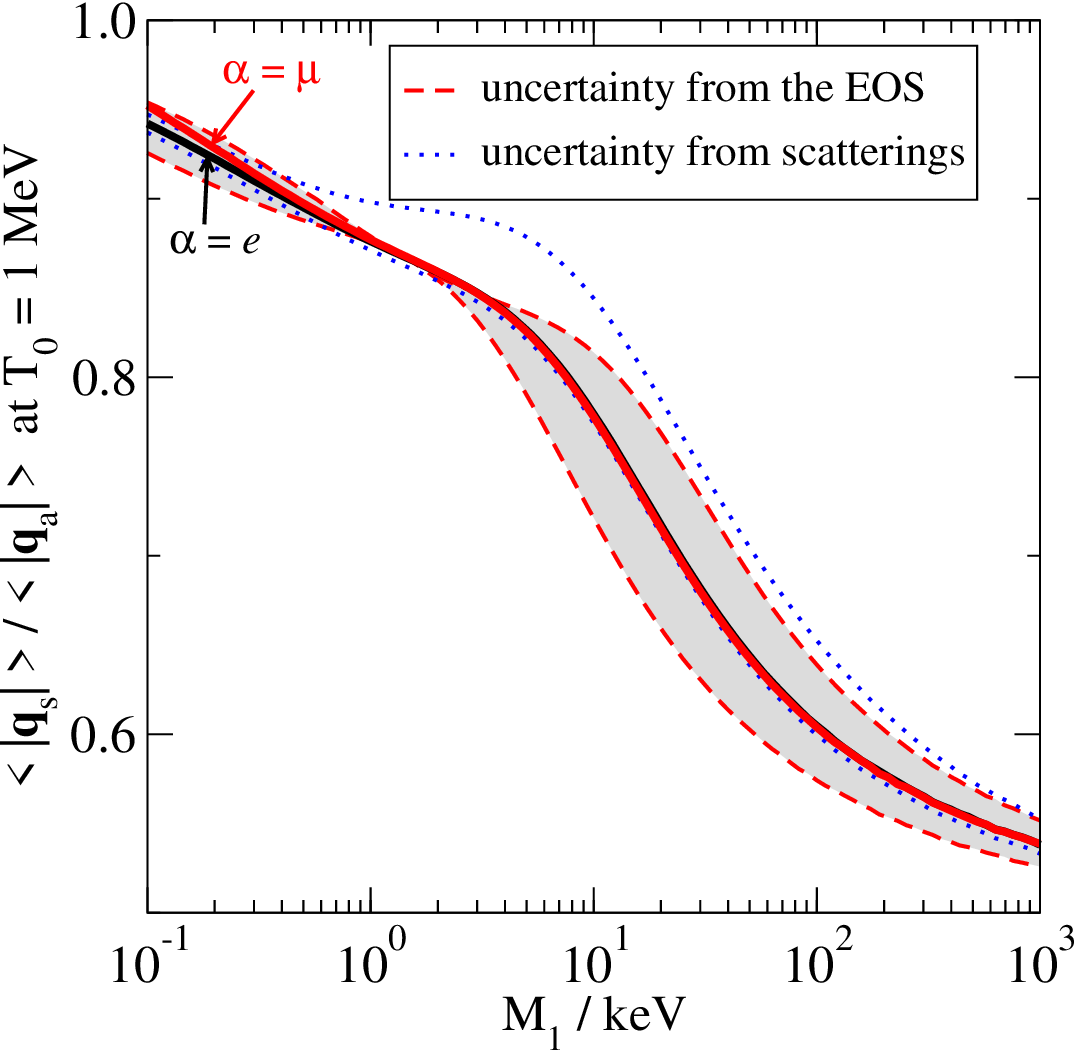}%
~~~\epsfysize=7.5cm\epsfbox{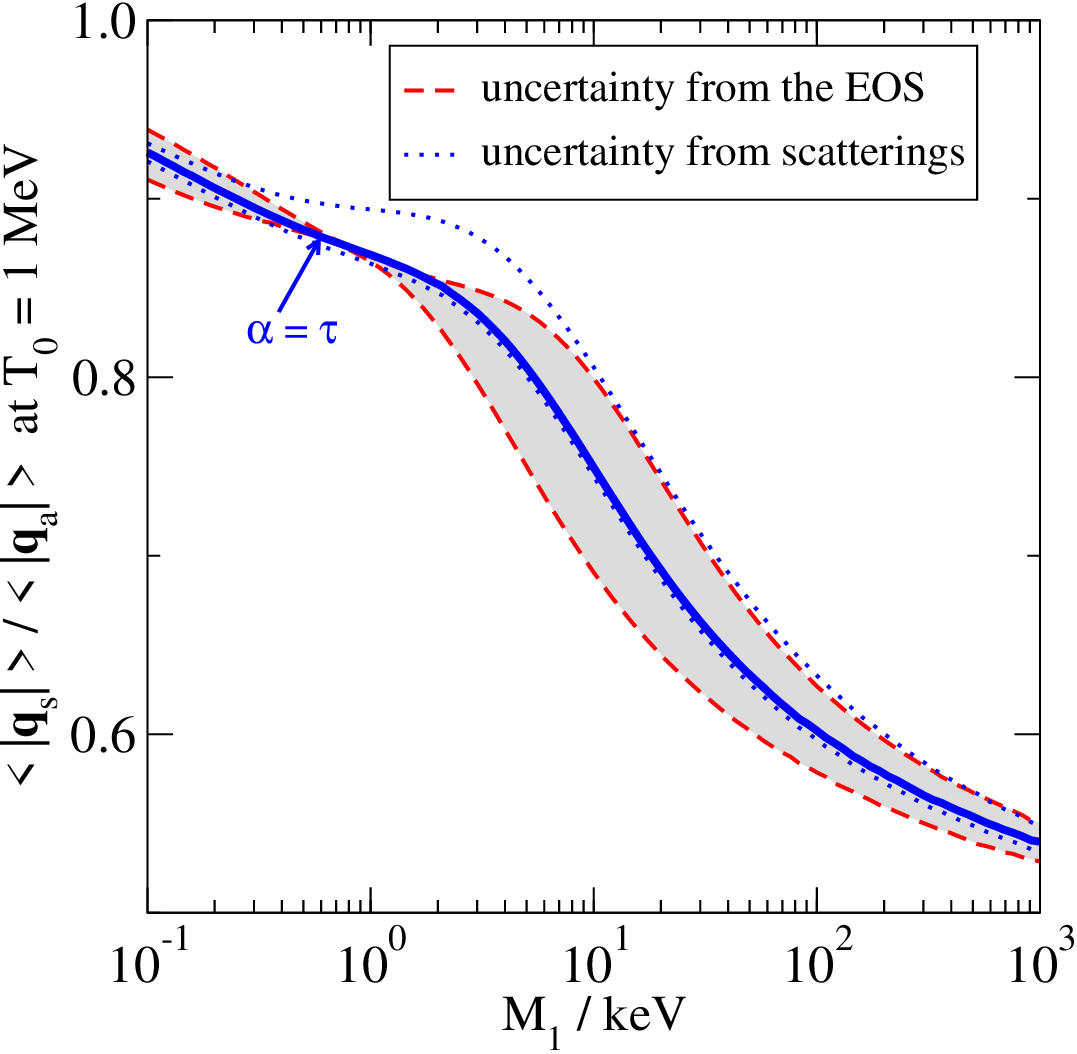}%
}

\caption[a]{\small  The average sterile neutrino 
momentum compared with the active neutrino 
equilibrium value, 
$
 \langle|\vec{q}_\rmi{s}|\rangle / \langle |\vec{q}_\rmi{a}| \rangle
$, 
where 
$
 \langle |\vec{q}_\rmi{a}| \rangle \equiv 7 \pi^4 T_0/180 \zeta(3) 
 \approx 3.15T_0
$. 
Left: $\alpha = e,\mu$, and only the uncertainties for $\alpha = e$ 
are shown.
Right: $\alpha = \tau$.
We have assumed for this figure that only one of 
the mixing angles $\theta_{\alpha 1}$ is non-vanishing.} 
\la{fig:avq}
\end{figure}

Finally, in \fig\ref{fig:distr} we show the momentum 
distribution, $n_{\alpha 1}(T,q)$, for a few 
examples, normalised to the massless equilibrium 
case, $n_\rmi{eq}(t_0,q) = 2 \nF{}(q) /(2\pi)^3$.
As noted in the literature~\cite{Abazajian:2005gj}, the momentum  
distribution is shifted towards the infrared compared  with the 
equilibrium distribution. For heavy $M_1$ the shift is fairly 
substantial (note the logarithmic scale). Structure formation 
simulations necessitate 
the momentum distribution function of dark matter as input, 
and curves such as the ones in \fig\ref{fig:distr} could be  
used for this purpose. However, as mentioned in \se\ref{se:intro}, 
one may get a rough first impression of the results already by 
just rescaling the average momentum to a shifted value. To allow
for such an estimate, \fig\ref{fig:avq} displays 
the average momentum of the  sterile neutrinos generated, 
in comparison with the equilibrium value for active neutrinos.  
\footnotetext[2]{
  Ref.~\cite{Abazajian:2005gj} does not state explicitly 
  the sterile neutrino mass range in which its \eq(7) should be 
  valid. After the eprint version of our paper had appeared, 
  we were informed by  K.~Abazajian that \eq(7) 
  of ref.~\cite{Abazajian:2005gj} was meant to be 
  valid for 0.5~keV $ < M_1 < $ 10~keV.
 }  

%
\section{The Dodelson-Widrow scenario}
\la{se:DW}

  We are now in the position to discuss the Dodelson-Widrow (DW)
  scenario for sterile neutrino dark matter, which is based on the
  assumptions (i)--(iv) we stated in \se\ref{se:intro}.  In this scenario, 
  the initial abundance of the dark-matter sterile neutrinos was zero 
  at $T \gg 1$ GeV, and they were only produced in active-sterile neutrino 
  transitions from the thermal plasma.
  Thus, in terms of \eq\nr{main}, it is the maximal value of the mixing 
  angle which realizes this scenario, $\sin^2(2\theta) =
  f(M_1)$.  We denote $\Ms \equiv M_1$ in this case.  
  This scenario can be confronted with a number of cosmological
  and astrophysical observations.

\begin{figure}[t]


\centerline{%
\epsfysize=9.0cm\epsfbox{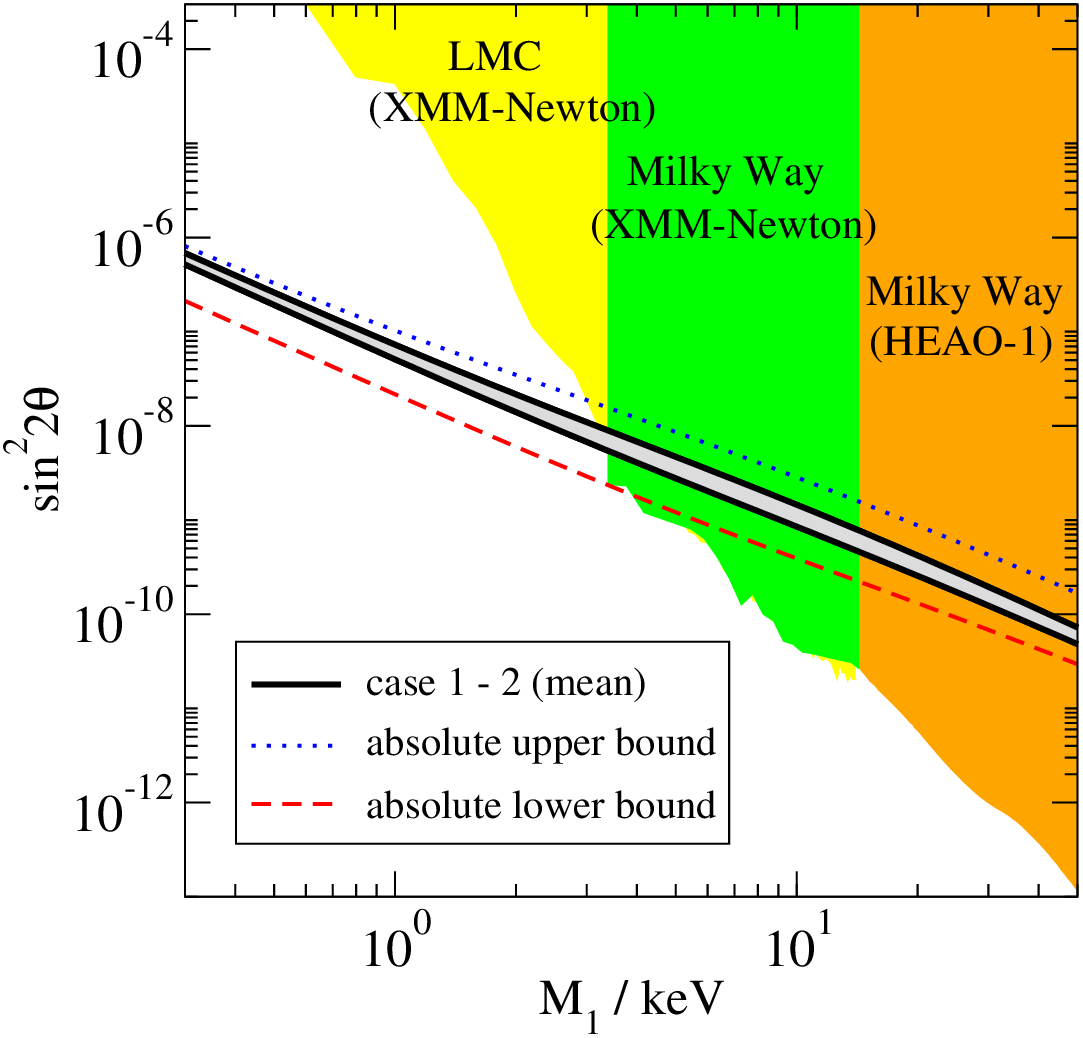}%
}

\caption[a]{\small 
The central region of \fig\ref{fig:exclusion_th}
($M_1/\mbox{keV} = 0.3 .... 50.0$), compared with
regions excluded by various X-ray constraints. From left to right, 
the X-ray constraints come from refs.~\cite{Boyarsky:2006fg}, 
\cite{Boyarsky:2006ag}, \cite{Boyarsky:2006fg}, respectively. 
The abbreviation LMC stands for the Large Magellanic Cloud, and 
the names in the parentheses refer to the X-ray observatories
whose data were employed for deriving the constraints.
} 
\la{fig:exclusion}
\end{figure}

In order to carry out a comparison with the various X-ray constaints 
that exist, we replot the curves of~\fig\ref{fig:exclusion_th} 
in~\fig\ref{fig:exclusion} (for a subrange of mass values), 
and compare with the regions excluded according to 
refs.~\cite{Boyarsky:2006fg}, \cite{Boyarsky:2006ag}.
To facilitate numerical estimates, we also note that in this 
mass range (and on the resolution of the logarithmic scale), the 
theoretical curves can be roughly approximated by straight lines, 
\be
 \begin{array}{rclcl}
  \log_{10}[\sin^2 \! 2\theta] & \simeq &  
  -6.9267  -  1.7350\;\log_{10}[\Ms/\mbox{keV}]
  & \hspace*{0.4cm} & \mbox{(absolute upper bound)} \\
  \log_{10}[\sin^2 \! 2\theta] & \simeq &  
  -7.0862  -  1.8451\;\log_{10}[\Ms/\mbox{keV}]
  & \hspace*{0.4cm} & \mbox{(case 2, mean)} \\
  \log_{10}[\sin^2 \! 2\theta] & \simeq &  
  -7.2491  -  1.8359\;\log_{10}[\Ms/\mbox{keV}]
  & \hspace*{0.4cm} & \mbox{(case 1, mean)} \\
  \log_{10}[\sin^2 \! 2\theta] & \simeq &  
  -7.6195  -  1.7428\;\log_{10}[\Ms/\mbox{keV}]
  & \hspace*{0.4cm} & \mbox{(absolute lower bound)}
 \end{array}
 \;. \la{fits}
\ee
These fits improve on the
first approximation in \eq\nr{constr1}. The lower bound of the 
region excluded by X-ray constaints will be denoted by $f_X(\Ms)$;
for this data a straight line is a worse approximation, 
but restricting to the mass range $M_s = (2 ... 10)$~keV,
the function can be roughly approximated by a parabola, 
\be
  \log_{10}[\sin^2 \! 2\theta] \simeq   
  -3.9499  -  9.7859\;\log_{10}[\Ms/\mbox{keV}]
  + 3.4902 \;\log_{10}^2[\Ms/\mbox{keV}]
 \;. \la{Xfit}
\ee

We then note that the intersection of $f(\Ms)$ and $f_X(\Ms)$ gives an
upper bound on $\Ms$.  It is found from 
\eqs\nr{fits}, \nr{Xfit} that the bound $\Ms < 3.5$ keV is obtained by
using the mean predictions for the cases 1 and 2 discussed in 
\eqs\nr{case1}, \nr{case2} of the
previous section.  For the most conservative case with respect to the
hadronic uncertainties,  
the bound can be as large as $\Ms < 4.3$ keV.
Adopting the absolute lower bound on $f(\Ms)$ with
the maximal hadronic uncertainties, and also assuming that 
the X-ray constraints are still uncertain 
by a factor of two,
we set $M_s < 6$~keV as the most conservative limit.
As we will show below, possible entropy dilution through the decays of
the heavier sterile neutrino(s) makes this bound more severe, and,
therefore, this bound is robust.
 
The fate of this scenario now depends on constraints from structure
formation simulations, making use of Lyman-$\alpha$ observations,
which give a lower bound on the mass of the dark matter sterile
neutrino as $M_s \gsim \langle |{\vec q}_{\rm s}| \rangle/
\langle |{\vec q}_{\rm a}| \rangle M_0$,
where $M_0 \approx 14.4$ keV (with $\langle |{\vec q}_{\rm s}| \rangle/
\langle |{\vec q}_{\rm a}| \rangle \simeq 0.9$)~\cite{Seljak:2006qw}
and $M_0 \approx 10$ keV
(with $\langle |{\vec q}_{\rm s}| \rangle/
\langle |{\vec q}_{\rm a}| \rangle \simeq 1.0$)~\cite{Viel:2006kd} 
at 95\% CL.
Note, however, that these bounds are
obtained by assuming that the sterile neutrino momentum distribution
function is proportional to the Fermi-Dirac one (with a certain
rescaling of the average momentum in ref.~\cite{Seljak:2006qw}), even
though the sterile neutrino is not thermalised.  As shown in
\fig\ref{fig:distr}, the momentum distribution function is in fact
significantly distorted from the Fermi-Dirac shape towards the
infrared (apart from having a smaller normalization). Assuming
that the corresponding effects can be captured by the
factor 
$\sim \langle |{\vec q}_{\rm s}| \rangle/
  \langle |{\vec q}_{\rm a} | \rangle$, we read from 
Fig.~\ref{fig:avq} that 
$\langle |{\vec q}_{\rm s}| \rangle/
\langle |{\vec q}_{\rm a} | \rangle \simeq 0.8$ 
for $\Ms \simeq 10$ keV. Given that
refs.~\cite{Seljak:2006qw,Viel:2006kd} assumed 
$\langle |{\vec q}_{\rm s}| \rangle/
\langle |{\vec q}_{\rm a} | \rangle \simeq 0.9,
1.0$, respectively, the results for lower bounds can be re-interpreted
as $\Ms \gsim 11.6$ keV and $\Ms \gsim 8$ keV.  Thus the X-ray
and Lyman-$\alpha$ regions do not overlap, and astrophysical and
cosmological constraints appear to rule out the DW scenario, in spite
of all theoretical uncertainties involved in the computation.  The
scenario can only survive if the computations of
refs.~\cite{Seljak:2006qw,Viel:2006kd} turn out to be uncertain for some
reason, be it on the observational side, or
on the theoretical side, related to the momentum distribution
of the sterile neutrino (i.e., the need to use of the proper 
distribution function from
\fig\ref{fig:distr} rather than the Fermi-Dirac one).
We should mention that, even in such a situation, the Tremaine-Gunn
lower bound $\Ms \gsim 0.3$ keV for fermionic dark matter does remain
valid~\cite{Dalcanton:2000hn}.
  
%
\subsection{Entropy dilution}

In the $\nu$MSM, there are two heavier sterile neutrinos 
in addition to the lightest dark-matter one.  As shown 
in ref.~\cite{Asaka:2006ek}, the heavier sterile neutrino(s) could
momentarily dominate the energy density of the Universe, and their
subsequent decay(s) could cause a significant entropy dilution 
after the dark matter sterile neutrinos have been produced. 
Here we briefly reiterate the corresponding effects, 
i.e.,\ relax the assumption (iv) of \se\ref{se:intro}.

Due to the entropy release,
the yield  $Y_1$ of the dark matter sterile neutrino is 
diluted by a factor $S$ compared with that without 
the entropy dilution. Thus, the upper bound on the 
active-sterile neutrino mixing angle becomes weaker,
\begin{eqnarray}
  \sin^2 (2 \theta) \lsim f(\Ms) \times {S} \,.
\end{eqnarray}
Note that $f_X(\Ms)$ from the X-ray constraints does not change,
since $f_X$ is derived from the present dark matter density 
$\Omega_{\rm dm}$.  Then, we can see from Fig.~\ref{fig:exclusion} that 
the upper bound on $\Ms$ becomes more stringent for a larger $S$.

On the other hand, the lower bound on
$\Ms$ from the Lyman-$\alpha$ observations also 
decreases if $S > 1$, by a factor $S^{1/3}$ if $N_1$ is
out-of-equilibrium~\cite{Asaka:2006ek}.
The combination of the two effects might open a window
for sterile neutrino dark matter even in the presence
of the Lyman-$\alpha$ constraints. To quantify
the effect, we note that in the region $\Ms\le 4$~keV, 
the X-ray bound can be approximated as 
$
  \log_{10}[\sin^2 \! 2\theta] \lsim   
  -4.6121  -  6.4659\;\log_{10}[\Ms/\mbox{keV}] 
$;
the most conservative bound from \eq\nr{fits} becomes
$
  \log_{10}[\sin^2 \! 2\theta] \lsim
  -7.6195  -  1.7428\;\log_{10}[\Ms/\mbox{keV}] + \log_{10}[S]
$;
and the Lyman-$\alpha$ bounds obtain the form
$\Ms\times S^{1/3} \gsim (8...11.6)$~keV.

If $\Ms\times S^{1/3} \gsim 8$~keV from ref.~\cite{Viel:2006kd}, 
we find that there appears an allowed region for $S \gsim 155$.  
When $S \approx 155$, a sterile neutrino with 
$\Ms \approx 1.5$~keV 
and $\sin^2 (2 \theta) \approx 1.9 \times 10^{-6}$ 
could function 
as dark matter and be consistent with X-ray and Lyman-$\alpha$ 
constraints. On the other hand, 
if $\Ms\times S^{1/3} \gsim 11.6$~keV
from ref.~\cite{Seljak:2006qw}, there is an allowed region for 
$S \gsim 3.3 \times 10^3$, corresponding 
to $\Ms \lsim 0.78$ keV 
and $\sin^2 (2 \theta) \gsim 1.2\times 10^{-4}$.
However, the combination
$
 ( {|M_D|_{\alpha 1}} / {0.1\;\mbox{eV}})^2
 = 2.5 \times 10^7 \; (\Ms / \mbox{keV})^2 \; \sin^2(2\theta)
$
is larger than $10^3$ in this case and, according to \fig\ref{fig:noneq}, 
$N_1$ are thermalized, whereby the assumptions we have made
are no longer self-consistent. We discuss the correct 
procedure for the latter case in~\se\ref{suse:thermal}.

To summarise, we have learned that the Dodelson-Widrow 
scenario remains a possibility
if there exists a large enough entropy dilution and if the lower
bound on $\Ms$ from the Lyman-$\alpha$ data is small enough.  However, 
it is very difficult to get such a large dilution factor within the $\nu$MSM.
It has been shown in ref.~\cite{Asaka:2006ek} that the $\nu$MSM
induces $S \sim 30$ at most and $S \gsim 100$ is obtained only if
there exists some physics beyond the $\nu$MSM.  Therefore, to realize
this scenario, the constraints from the Lyman-$\alpha$ data should again 
be weaker than those presented in refs.~\cite{Seljak:2006qw,Viel:2006kd}.

%
\subsection{The case of thermalised sterile neutrinos}
\la{suse:thermal}

So far, we have assumed that the Dirac neutrino masses 
$|M_D|_{\alpha 1}$ are sufficiently small so that $N_1$ was never 
in thermal equilibrium.  Now, let us discuss the possibility 
of thermalized $N_1$ as dark matter.  
If $N_1$ is thermalized at some high temperatures 
and decouples at $T = T_\rmi{d1}$,  
the present relic density is given by
\begin{eqnarray}
  \Omega_{N_1} h^2 = 10.6 \,
  \left( \frac{\Ms}{\rm keV} \right)
  \left( \frac{10.75}{h_{\rm eff}(T_\rmi{d1})} \right) \,.
\end{eqnarray}
Note that this result is independent of 
$\theta$ as long as $\theta$ is sufficiently 
large such that $N_1$ was thermalized at $T > T_\rmi{d1}$.
To explain all of dark matter by such a density of 
$N_1$'s, the sterile neutrino mass should be
\begin{eqnarray}
  \Ms \simeq 10 \,\mbox{eV} 
  \left( \frac{h_{\rm eff}(T_\rmi{d1})}{10.75} \right) \,.
\end{eqnarray}
The ratio $h_\rmi{eff}/10.75$ is at least unity, 
cf.\ \fig\ref{fig:eweos}. 
In any case, this value is much smaller than the Tremaine-Gunn bound, 
$\Ms \gsim 0.3$ keV, and is also below the bounds from the Lyman-$\alpha$ 
constraints for the thermalised case; 
$\Ms \gsim 2.4$ keV~\cite{Seljak:2006qw} and
$\Ms \gsim 2$ keV~\cite{Viel:2006kd} (at 95\% CL).

If there exists entropy dilution, 
the mass of the once-thermalized dark matter sterile neutrino becomes 
\begin{eqnarray}
  \Ms \simeq 10 \,\mbox{eV} 
  \left( \frac{h_{\rm eff}(T_\rmi{d1})}{10.75} \right) 
  \times S \,.
\end{eqnarray}
On the other hand, the Tremaine-Gunn bound and the Lyman-$\alpha$ 
constraint remain unchanged. This is because the distribution 
function of the sterile neutrino was the Fermi-Dirac one 
at $T = T_\rmi{d1}$ and it is just red-shifted to the present time.
Although the Tremaine-Gunn bound can be satisfied with 
$S \simeq 30$, larger factors $S \gsim 240$ and $200$ 
are required to avoid the Lyman-$\alpha$ bounds from 
refs.~\cite{Seljak:2006qw,Viel:2006kd}, respectively.  Therefore, 
the thermalized $N_1$ can be dark matter within the $\nu$MSM only
if there is the largest possible entropy dilution and if the lower 
bound on $\Ms$ is given by the Tremaine-Gunn bound, rather than 
the Lyman-$\alpha$ bounds.

%
\subsection{Summary}

We have shown in this section that if the lower bound on $\Ms$ 
from the Lyman-$\alpha$ data is indeed as found 
in refs.~\cite{Seljak:2006qw,Viel:2006kd}, then there is no parameter 
space left within the $\nu$MSM to realize the scenario of 
sterile neutrinos serving as warm dark matter, respecting
the ``crucial'' assumptions (i)--(iii) of \se\ref{se:intro}.

Of course, this does not mean that the sterile neutrino is excluded
as a dark matter candidate, but that at least one of 
the assumptions would have to be relaxed for this to be the case. 
For instance, sterile neutrinos 
could be produced mainly in interactions beyond the $\nu$MSM, 
as in ref.~\cite{Shaposhnikov:2006xi}. Another logical possibility 
is that the Universe had substantial leptonic asymmetries at small 
temperatures, leading to a resonant production of sterile 
neutrinos~\cite{Shi:1998km}. In any event, 
for $\Ms > 3.5$ keV (6 keV if all hadronic uncertainties are pushed
in one direction and the strongest X-ray bounds are relaxed by a factor 
of two) the 
active-sterile transitions from the thermal charge-symmetric plasma cannot 
produce cosmologically interesting amounts of sterile neutrino dark matter. 

  Finally, we should like to stress that, 
  within the corner of the parameter space that defines the $\nu$MSM, 
  the bound on the mixing angle $\sin^2 (2 \theta) \lsim f(\Ms)$,  
  i.e.\ Eq.(\ref{constr0}), gives a prediction 
  on the mass scale of the lightest active
  neutrino, $m_{\nu_1}$, whose see-saw formula includes
  the lightest Majorana mass $M_1$ only~\cite{Asaka:2005an}.  
  By setting $C_\alpha (M_1) \simeq 0.5$, we find
  \begin{eqnarray}
    m_{\nu_1} \lsim
    \sum_{\alpha = e, \mu, \tau} \frac{|M_D|_{\alpha 1}^2}{M_1}
    \lsim
    2 \times 10^{-5} \; \mbox{eV} 
    \left( \frac{ \mbox{keV} }{ M_1 } \right) \,.
  \end{eqnarray}
  If $N_1$ is the dark matter particle, 
  its mass should in any case be $M_1 \gsim 0.3$
  keV from the Tremaine-Gunn bound.  
  Then, we get the weakest bound 
  as $m_{\nu_1} \lsim 6.7 \times 10^{-5}$ eV, which is much smaller 
  than the neutrino mass scales observed in the atmospheric and solar
  neutrino experiments.  Furthermore, the upper bound on $m_{\nu_1}$
  becomes smaller for a larger $M_1$.
  This clearly excludes the possibility
  that the three active neutrinos are nearly degenerate in mass, 
  and indicates that one could find  
  the absolute values of the heavier active neutrino masses
  from the results of neutrino oscillation experiments~\cite{Asaka:2005an}.
  These conclusions remain valid even if there is entropy production
  as long as non-thermalized sterile neutrinos play the role of 
  dark matter.

%
\section{Conclusions}
\la{se:concl}
In this paper we have studied the production of sterile neutrinos 
in the $\nu$MSM through active-sterile neutrino transitions from 
a thermal plasma, which does not contain any significant asymmetries
related to lepton numbers. Though a first-principles formula for
the production rate exists, which is valid to all orders in the 
strong coupling constant, its practical evaluation is subject 
to a number of uncertainties related to the strong interactions. 
The purpose of this paper has been to present a realistic
``mean'' evaluation of the hadronic contributions, 
and to estimate conservatively
all the uncertainties that this mean result is subject to.

The most important 
result of this paper is encoded in the four lines
shown in \fig\ref{fig:exclusion}. They correspond to the case when 
there is no entropy production ($S=1$) due to the decay of the heavier 
sterile neutrinos of the $\nu$MSM. The area above the dotted line is 
certainly excluded: the amount of dark matter produced would lead to
the overclosure of the Universe. The region below the dashed line is
certainly allowed: the amount of sterile neutrinos produced due to
active-sterile transitions is smaller than the amount of dark matter
observed. Any point in the region between the two solid lines
(corresponding to the ``most reasonable'' model for the hadronic
contributions that we have been able to come up with) 
can lead to dark matter generation entirely due to
active-sterile transitions.  A maximal variation of the parameters
of our hadronic model extends this region to the space between 
the dotted and dashed lines. In the case of entropy production with 
a factor $S>1.0$, all these four lines simply move up by a factor $S$.

As \fig\ref{fig:exclusion} shows, 
active-sterile transitions can account for all of dark matter only if
$M_1 < 3.5$ keV, if the ``most reasonable'' hadronic model is taken.
The most conservative upper limit would correspond to  
$M_1 < 6$ keV,
if all uncertainties are pushed in the same direction
and also if the most stringent X-ray bounds are relaxed by 
a factor of two. 
Therefore, if the Lyman-$\alpha$ constraints from
refs.~\cite{Seljak:2006qw,Viel:2006kd} are taken for granted, the
production of sterile neutrinos due to active-sterile neutrino
transitions happens to be too small to account for the observed
abundance of dark matter. In other words, physics beyond the $\nu$MSM
is likely to be required to produce dark matter sterile neutrinos.
Another option is to assume that the Universe contained relatively
large lepton asymmetries~\cite{Shi:1998km}.  We would like to stress,
though, that (apart from the astrophysical uncertainties related to
the Lyman-$\alpha$ data) the simulations mentioned have not utilised
the correct non-equilibrium momentum distribution functions as given
in \fig\ref{fig:distr}, and may thus contain systematic uncertainties.

%
\section*{Acknowledgements}

The work of T.A.\ was supported in part by the grants-in-aid from 
the Ministry of Education, Science, Sports, and Culture of Japan,
Nos.\ 16081202, 17340062 and 18740122,  and that of M.S.\ by the Swiss National
Science Foundation.

\newpage


\appendix
\renewcommand{\thesection}{Appendix~\Alph{section}}
\renewcommand{\thesubsection}{\Alph{section}.\arabic{subsection}}
\renewcommand{\theequation}{\Alph{section}.\arabic{equation}}

%
\section{Two-particle phase space integrals}

We discuss in this appendix how the phase space integrals
of \eq\nr{pert1} can be reduced to a one-dimensional integration
in order to allow for a numerical evaluation. 

Let $P$ be a time-like four-vector, $P^2 > 0$. We can then define
a Lorentz-boost, $\Lambda_P$, which takes us to the rest-frame
with respect to $P$: $[\Lambda_P P]^i = 0$, $i = 1,2,3$. 
We denote $\hat Q \equiv \Lambda_P Q$, with the components 
$\hat Q = (\hat q^0,\hat\vec{q})$. To be explicit, this Lorentz-boost
is given by
\ba
 \hat q^0 & = & \gamma [q^0 + \beta \vec{e}\cdot \vec{q}]
 \;, \\
 \hat\vec{q} & = & 
 \vec{q} - \vec{e}\cdot\vec{q} \, \vec{e} + 
 \gamma [\beta q^0 + \vec{e}\cdot \vec{q}] \vec{e} 
 \;,  
\ea
where
\be
 \beta = -\frac{|\vec{p}|}{p^0}
 \;, \quad
 \gamma = \frac{1}{\sqrt{1 - \beta^2}}
 \;, \quad
 \vec{e} = \frac{\vec{p}}{|\vec{p}|}
 \;.
\ee

Denoting by $f[u;P_1;P_2;Q]$ a generic Lorentz-scalar, for instance
\be
 f[u;P_1;P_2;Q] = 
 2\, Q \cdot P_1 \;
 \nB{}(u \cdot P_2)[1-\nF{}(u \cdot P_1)] 
 \;, \la{bex}
\ee
as would be relevant for the second term in \eq\nr{pert1},
we then consider the integral
\be
   I
 \equiv 
  \int \! \frac{{\rm d}^3 \vec{p}_1}{(2\pi)^3 2 E_1} \,
  \int \! \frac{{\rm d}^3 \vec{p}_2}{(2\pi)^3 2 E_2} \,
  (2\pi)^4 \delta^{(4)}(P_2-P_1-Q) \,
 f[u;P_1;P_2;Q]
 \;,
 \la{pre_example}
\ee
where $Q= (q^0,\vec{q})$, $P_i=(E_i,\vec{p}_i)$ are on-shell four-vectors.
Making use of the Lorentz-boost introduced above, 
we rewrite the integration
in a frame where $Q$ is at rest, 
$\hat \vec{q} = \vec{0}$: 
\be
   I
  = 
  \int \! \frac{{\rm d}^3 \hat\vec{p}_1}{(2\pi)^3 2 \hat E_1} \,
  \int \! \frac{{\rm d}^3 \hat\vec{p}_2}{(2\pi)^3 2 \hat E_2} \,
  \left.
  (2\pi)^4 \delta^{(3)}(\hat\vec{p}_2 - \hat\vec{p}_1) 
  \delta (\hat E_2-\hat E_1-M_I) \,
 f[\hat u;\hat P_1;\hat P_2;\hat Q]
 \right|_{\hat V \equiv \Lambda_{Q} V}
 \;.
\ee
The integral over $\hat\vec{p}_2$ is now trivial: 
\ba
 & &  \hspace*{-1.8cm}
   I
  = 
  \frac{1}{(2\pi)^2}
  \int \! \frac{{\rm d}^3 \hat\vec{p}_1}
  {4 \sqrt{\hat\vec{p}_1^2 + m_1^2} \sqrt{\hat\vec{p}_1^2 + m_2^2} } \,
  \delta( \sqrt{\hat\vec{p}_1^2 + m_2^2} - \sqrt{\hat\vec{p}_1^2 + m_1^2} 
  - M_I ) \,
   \times   \nn 
  & \times & \!\!\! 
 f[\Lambda_{Q} u;
  (\sqrt{\hat\vec{p}_1^2 + m_1^2},
  |\hat\vec{p}_1|\Omega_{\hat\vec{p}_1});
  (\sqrt{\hat\vec{p}_1^2 + m_2^2},
   |\hat\vec{p}_1|\Omega_{\hat\vec{p}_1});
  \Lambda_{Q} Q]
 \biggr\}
 \;. 
\ea
Moreover the integral over $|\hat\vec{p}_1|$ can be performed by making
use of 
\be
 \int_0^\infty \! 
 \frac{{\rm d}x\, x^2 g(x)}{\sqrt{x^2 + m_1^2}\sqrt{x^2 + m_2^2}}
 \delta(\sqrt{x^2 + m_2^2} - \sqrt{x^2 + m_1^2} - M)
 = 
 \theta(m_2 - m_1 - M) 
 \frac{\rho_{12}(M) g(\rho_{12}(M))}{M} 
 \;,
\ee
where the function
\be
 \rho_{ij}(M) \equiv \frac{1}{2 M} 
 \sqrt{M^4 - 2 M^2 (m_i^2 + m_j^2) + 
 (m_i^2 - m_j^2)^2}
 \la{rhoij}
\ee
is real and positive for $0 < M < |m_1 - m_2|$ and 
$M > m_i + m_j$. The integral thus becomes
\ba
 & &  \hspace*{-1.5cm}
   I
  = 
  \frac{|\hat\vec{p}_1|}{(4\pi)^2 M_I}
  \theta(m_2 - m_1 - M_I)  
   \times   \la{pre_fin} \\
  & & \hspace*{-1.2cm} \times 
 \left.
 \int \! {\rm d}\Omega_{\hat\vec{p}_1}\, 
 f[
  \Lambda_{Q} u;
  (\sqrt{\hat\vec{p}_1^2 + m_1^2},
  |\hat\vec{p}_1|\Omega_{\hat\vec{p}_1});
  (\sqrt{\hat\vec{p}_1^2 + m_2^2},
  |\hat\vec{p}_1|\Omega_{\hat\vec{p}_1});
  (M_I,\vec{0})
  ]
 \right|_{|\hat\vec{p}_1| \equiv \rho_{12}(M_I)}
 \;. \nonumber
\ea

Now, the integrand in~\eq\nr{pre_fin} depends on two spatial vectors, 
$\vec{q}$, $\Omega_{\hat\vec{p}_1}$, and thus on one angle. 
We can choose 
$
 \vec{q} = (0,0,|\vec{q}|)
$, 
$
 \Omega_{\hat\vec{p}_1} = (\sin\theta,0,\cos\theta)
$, 
$
 \int \! {\rm d}\Omega_{\hat\vec{p}_1} = 2\pi 
 \int_0^\pi \! {\rm d}\theta \, \sin\theta
$.
However, the Fermi-distributions in \eq\nr{bex}
do depend on $\theta$, because the spatial part of 
$\Lambda_Q u$ is non-zero and proportional to $\vec{q}$, 
so that the remaining
integration is non-trivial. On the contrary, scalar products 
such as $Q\cdot P_1$, $Q\cdot P_2$, $P_1\cdot P_2$ are independent
of $\theta$, 
because we can evaluate them in the ``hatted'' frame
where $\hat\vec{q} = \vec{0}$. 

%
\section{Three-particle phase space integrals}

We discuss in this appendix how the phase space integrals
of \eq\nr{pertF} can be reduced to a three-dimensional integration
in order to allow for a numerical evaluation. 

Denoting by $f[u;P_1;P_2;P_3;Q]$ a generic Lorentz-scalar, for instance
\be
 f[u;P_1;P_2;P_3;Q] = 
 \mathcal{T}_i \,
 \nF{}(u\cdot P_1)\nF{}(u \cdot P_2)[1-\nF{}(u \cdot P_3)] 
 \;, \la{fex}
\ee
where $\mathcal{T}_i = \mathcal{T}_1$ or $\mathcal{T}_2$
from \eqs\nr{T1} or \nr{T2}, 
we consider the integral
\be
   I
 \equiv 
  \int \! \frac{{\rm d}^3 \vec{p}_1}{(2\pi)^3 2 E_1} \,
  \int \! \frac{{\rm d}^3 \vec{p}_2}{(2\pi)^3 2 E_2} \,
  \int \! \frac{{\rm d}^3 \vec{p}_3}{(2\pi)^3 2 E_3} \,
  (2\pi)^4 \delta^{(4)}(P_1+P_2-P_3-Q) \,
 f[u;P_1;P_2;P_3;Q]
 \;,
 \la{example}
\ee
where $Q= (q^0,\vec{q})$, $P_i=(E_i,\vec{p}_i)$ are on-shell four-vectors.
Making use of the Lorentz-boost introduced in Appendix~A, 
we rewrite a certain Lorentz-invariant subpart of the integration
in a frame where $P_3 + Q$ is at rest, 
$\hat \vec{p}_3 + \hat \vec{q} = \vec{0}$: 
\ba
 & &  \hspace*{-1.8cm}
   I
  = 
  \int \! \frac{{\rm d}^3 \vec{p}_3}{(2\pi)^3 2 E_3} \, \biggl\{ 
  \int \! \frac{{\rm d}^3 \hat\vec{p}_1}{(2\pi)^3 2 \hat E_1} \,
  \int \! \frac{{\rm d}^3 \hat\vec{p}_2}{(2\pi)^3 2 \hat E_2} \,
  \times  
  \nn 
  & \times & \!\!\!
  (2\pi)^4 \delta^{(3)}(\hat\vec{p}_1 + \hat\vec{p}_2) 
  \delta (\hat E_1+\hat E_2-\hat E_3-\hat q^0) \,
 f[\hat u;\hat P_1;\hat P_2;\hat P_3;\hat Q]
 \biggr\}_{\hat V \equiv \Lambda_{P_3 + Q} V}
 \;.
\ea
The integral over $\hat\vec{p}_2$ is now trivial: 
\ba
 & &  \hspace*{-1.8cm}
   I
  = 
  \int \! \frac{{\rm d}^3 \vec{p}_3}{(2\pi)^3 2 E_3} \, \biggl\{ 
  \frac{1}{(2\pi)^2}
  \int \! \frac{{\rm d}^3 \hat\vec{p}_1}
  {4 \sqrt{\hat\vec{p}_1^2 + m_1^2} \sqrt{\hat\vec{p}_1^2 + m_2^2} } \,
  \delta( \sqrt{\hat\vec{p}_1^2 + m_1^2} + \sqrt{\hat\vec{p}_1^2 + m_2^2} 
  - \hat E_3 - \hat q^0 ) \,
   \times   \nn 
  & \times & \!\!\!
 f[\Lambda_{P_3 + Q} u;
  (\sqrt{\hat\vec{p}_1^2 + m_1^2},
  |\hat\vec{p}_1|\Omega_{\hat\vec{p}_1});
  (\sqrt{\hat\vec{p}_1^2 + m_2^2},
   -|\hat\vec{p}_1|\Omega_{\hat\vec{p}_1});
  \Lambda_{P_3 + Q} P_3;
  \Lambda_{P_3 + Q} Q]
 \biggr\}
 \;. \nn 
\ea
Moreover the integral over $|\hat\vec{p}_1|$ can be performed by making
use of 
\be
 \int_0^\infty \! 
 \frac{{\rm d}x\, x^2 g(x)}{\sqrt{x^2 + m_1^2}\sqrt{x^2 + m_2^2}}
 \delta(\sqrt{x^2 + m_1^2} + \sqrt{x^2 + m_2^2} - \hat E)
 = 
 \theta(\hat E - m_1 - m_2) 
 \frac{\rho_{12}(\hat E) g(\rho_{12}(\hat E))}{\hat E} 
 \;,
\ee
where the function $\rho_{ij}$ is defined in \eq\nr{rhoij}. 
The integral thus becomes
\ba
 & &  \hspace*{-1.5cm}
   I
  = 
  \int \! \frac{{\rm d}^3 \vec{p}_3}{(2\pi)^3 2 E_3} \, \biggl\{ 
  \frac{1}{(4\pi)^2}
  \int \! {\rm d}\Omega_{\hat\vec{p}_1}\, 
  \theta(\hat E_3 + \hat q^0 - m_1 - m_2) 
  \frac{\rho_{12}(\hat E_3 + \hat q^0)}{\hat E_3 + \hat q^0}  \,
   \times   \nn 
  & & \hspace*{-1.5cm} \times 
 f[
  \Lambda_{P_3 + Q} u;
  (\sqrt{\hat\vec{p}_1^2 + m_1^2},
  |\hat\vec{p}_1|\Omega_{\hat\vec{p}_1});
  (\sqrt{\hat\vec{p}_1^2 + m_2^2},
 -|\hat\vec{p}_1|\Omega_{\hat\vec{p}_1});
  \Lambda_{P_3 + Q} P_3;
  \Lambda_{P_3 + Q} Q
  ]
 \biggr\} 
 \;, \la{fin} 
\ea
where $|\hat\vec{p}_1| \equiv \rho_{12}(\hat E_3 + \hat q^0)$.

Now, the integrand in~\eq\nr{fin} depends on three spatial vectors, 
$\vec{q}$, $\vec{p_3}$, $\Omega_{\hat\vec{p}_1}$, and thus 
in general on three angles. However, the Fermi-distributions
only depend on two angles, so that the dependence on the 
third angle is very simple and can be handled analytically.
To implement this in practice we may for instance note that, 
due to O(3) invariance, we can choose $\vec{q}\equiv (0,0,|\vec{q}|)$ 
and $\vec{p}_3$ in the $(x,z)$-plane,
$\vec{p}_3 \equiv |\vec{p}_3|(\sin\theta,0,\cos\theta)$, 
so that 
\be
 \int\! {\rm d}^3 \vec{p}_3 \equiv 
 2\pi \int_0^\infty \! |\vec{p}_3|^2 \, {\rm d}|\vec{p}_3| 
 \int_0^\pi \! \sin\theta \,{\rm d}\theta
 \;.
\ee 
Defining the unit vector 
$
 \vec{e} = (\vec{p}_3 + \vec{q})/|\vec{p}_3 + \vec{q}|
$
as before [it now lies within the ($x,z$)-plane], 
we parametrize the remaining vector $\Omega_{\hat\vec{p}_1}$
by using spherical coordinates $\hat\theta, \hat\phi$
with $\vec{e}$ as the polar axis. In the original frame, 
$\Omega_{\hat\vec{p}_1}$ is then given by 
$
 \Omega_{\hat\vec{p}_1} \equiv 
 (e_z \sin\hat\theta\cos\hat\phi + e_x \cos\hat\theta, ~
      \sin\hat\theta\sin\hat\phi, ~
  e_z \cos\hat\theta - e_x \sin\hat\theta\cos\hat\phi)
$.
With this parametrization and given the form of $f$ that 
appears in \eq\nr{fex}, 
it is not difficult to realise that the integrand
depends on $\hat\phi$ only as a 2nd order polynomial in $\cos\hat\phi$:
$f(\hat\phi) = a + b \cos\hat\phi + c \cos^2\!\hat\phi$. Therefore
we can replace
\be
  \int_0^{2\pi} \! {\rm d}\hat \phi \, f(\hat\phi)
 = \pi \Bigl[ 
  f\Bigl( \frac{\pi}{4} \Bigr) + 
  f\Bigl( \frac{3 \pi}{4} \Bigr)
  \Bigr]
  \;.
\ee
Only a three-dimensional integration (over $|\vec{p}_3|, \theta, \hat\theta$)
needs hence to be carried out. 

Finally we remark that in the other channels ($3\to 1$, $1\to 3$),
the role of $P_3 + Q$ is played a difference, for instance $P_3 - Q$.
For arbitrary $P_3,Q$ this is not necessarily time-like. However the 
$\delta$-functions appearing in these channels always restrict 
the differences to be equal to a sums, for instance $P_1 + P_2$, 
which are time-like. Therefore non-zero contributions only emerge from 
regions of the phase space where $P_3 - Q$ is time-like, and the 
procedure described above can be taken over with minimal modifications.

\newpage


\section*{Erratum}

\newcommand{\aL}{a^{ }_\rmii{L}}
\newcommand{\aR}{a^{ }_\rmii{R}}
\setcounter{equation}{0}
\makeatletter \@addtoreset{equation}{section} \makeatother
\renewcommand{\theequation}{\arabic{equation}}

\vspace*{-0.95cm}

\hfill (January 2015)

\vspace*{0.5cm}

The following corrections should be implemented in the formulae. 
In \eq(3.6), the two factors $\fr34$ should be replaced with $\fr12$.
In table~1, the factor $\fr94$ should be replaced with $\fr32$, and 
the factor $-\fr34$ with $-\fr12$.
In addition, in \eqs(3.7) and (3.8), one should 
switch $\aL\leftrightarrow\aR$, which implies $b\to -b$, $d\to -d$.
The numerical effect of these corrections on $\im\Sigma$
can be up to $\sim 25\%$ at $T < 20$~MeV for $\alpha = 1$, 
but is only on the few percent level
at $T > 100$~MeV. The effect on the right-handed neutrino
production rate is thus
much smaller than hadronic uncertainties at $T > 100$~MeV. 

The reason for the last correction is subtle. 
Given that weak interactions of Standard Model neutrinos 
take place through vertices with the Dirac structure $\sim \gamma^\mu \aL$, 
where $\aL \equiv (1-\gamma^{ }_5)/2$ is a chiral projector, 
their inverse propagator is of the form~[52] 
\be
 S^{-1}(Q) = \aR \, \bigl( \bsl{Q} + \bsl{\Sigma} \bigr) \, \aL
 \;. \la{invS}
\ee 
Because of the chiral projectors, the (retarded) 
self-energy can be expressed as
\be
 \bsl{\Sigma} = a\, \bsl{Q} + b\, \msl{u}
 \;, \la{structure}
\ee
where 
$a,b$ are complex functions. With this form, 
the left-handed neutrino propagator reads~[52]
\be
 S(Q) = \aL \, 
  \frac{ (1+a) \, \bsl{Q} + b\, \msl{u} }{ [(1+a)\, Q + b\, u ]^2}
  \, \aR
 \;. \la{S}
\ee
It is this propagator which plays a role in our expressions.
However, the imaginary parts
of $a,b$ originate at 2-loop level and then
$\bsl{\Sigma}$ has a chiral structure which is {\em not}
trivially of the form in \eq\nr{structure}. 
For instance, hadronic effects contain a phase space integral 
(cf.\ \eq(3.4))
\ba
 \im \bsl{\Sigma} & \propto & 
 \int \! {\rm d}\Omega^{ }_{2\to 2} \, 
 \; \gamma^{\mu} \bsl{P}^{ }_{\! 1} \gamma^\nu 
 \; \tr\Bigl[
  \bsl{P}^{ }_{\! 2}\,  \gamma_\mu\, 
  \bsl{P}^{ }_{\! 3}\,  \gamma_\nu\, \aL 
 \Bigr]
 \nn 
 & = & 
 8 \int \! {\rm d}\Omega^{ }_{2\to 2} \, 
 \Bigl( 
   P^{ }_1 \cdot P^{ }_2 \; \bsl{P}^{ }_{\! 3} \; \aR + 
   P^{ }_1 \cdot P^{ }_3 \; \bsl{P}^{ }_{\! 2} \; \aL
 \Bigr)
 \;, \la{Sigma} 
\ea 
where ${\rm d}\Omega^{ }_{2\to 2}$ denotes a phase space 
measure including appropriate initial or final state Fermi distributions, 
and $P^{ }_1, P^{ }_2$ and $P^{ }_3$ are four-momenta of the associated
particles. Now, 
when considered in the context of \eq\nr{invS}, the part containing $\aR$ 
gets projected out.
In contrast, if the full $\bsl{\Sigma}$ were inserted into
the numerator of \eq\nr{S} (which is inconsistent), the
part containing $\aL$ would get projected out. In order to get
correct results, it is important to {\em first} take the projection
that sets the self-energy in the form 
of \eqs\nr{invS}, \nr{structure}. 

\begin{itemize}
 \item[{[52]}]
  H.A.~Weldon,
  Phys.\ Rev.\ D {26} (1982) 2789.
\end{itemize}



\begin{thebibliography}{99}

%
\bibitem{Asaka:2005an}
  T.~Asaka, S.~Blanchet and M.~Shaposhnikov,
  Phys.\ Lett.\ B {631} (2005) 151 
  [hep-ph/0503065].

\bibitem{Asaka:2005pn}
  T.~Asaka and M.~Shaposhnikov,
  Phys.\ Lett.\ B {620} (2005) 17
  [hep-ph/0505013];
%
  M.~Shaposhnikov,
  Nucl.\ Phys.\ B {763} (2007) 49
  [hep-ph/0605047].

\bibitem{Akhmedov:1998qx}
  E.K.~Akhmedov, V.A.~Rubakov and A.Y.~Smirnov,
  Phys.\ Rev.\ Lett.\  {81} (1998) 1359
  [hep-ph/9803255].
  
\bibitem{Kuzmin:1985mm}
  V.A.~Kuzmin, V.A.~Rubakov and M.E.~Shaposhnikov,
  Phys.\ Lett.\ B {155} (1985) 36.
  
\bibitem{Shaposhnikov:2006xi}
  M.~Shaposhnikov and I.~Tkachev,
  Phys.\ Lett.\  B {639} (2006) 414
  [hep-ph/0604236].


\bibitem{astro}
  A.~Kusenko and G.~Segr\`e,
  Phys.\ Lett.\ B {396} (1997) 197
  [hep-ph/9701311];
%
  G.M.~Fuller, A.~Kusenko, I.~Mocioiu and S.~Pascoli,
  Phys.\ Rev.\ D {68} (2003) 103002
  [astro-ph/0307267];
%
  M.~Barkovich, J.C.~D'Olivo and R.~Montemayor,
  Phys.\ Rev.\ D {70} (2004) 043005
  [hep-ph/0402259];
%
  M.~Mapelli, A.~Ferrara and E.~Pierpaoli,
  Mon.\ Not.\ Roy.\ Astron.\ Soc.\  {369} (2006) 1719
  [astro-ph/0603237];
%
  E.~Ripamonti, M.~Mapelli and A.~Ferrara,
  astro-ph/0606482;
%
  E.~Ripamonti, M.~Mapelli and A.~Ferrara,
  astro-ph/0606483;
%
  P.L.~Biermann and A.~Kusenko,
  Phys.\ Rev.\ Lett.\  {96} (2006) 091301
  [astro-ph/0601004];
%
  J.~Stasielak, P.L.~Biermann and A.~Kusenko,
  astro-ph/0606435;
%
%
  A.~Kusenko,
  hep-ph/0609081; 
%
  F.~Munyaneza and P.L.~Biermann,
  astro-ph/0609388;
%
 J.~Hidaka and G.M.~Fuller,
  astro-ph/0609425.
  
\bibitem{Dodelson:1993je}
  S.~Dodelson and L.M.~Widrow,
  Phys.\ Rev.\ Lett.\  {72} (1994) 17 
  [hep-ph/9303287].

\bibitem{Shi:1998km}
  X.~Shi and G.M.~Fuller,
  Phys.\ Rev.\ Lett.\  {82} (1999) 2832
  [astro-ph/9810076].
  
\bibitem{Dolgov:2000ew}
  A.D.~Dolgov and S.H.~Hansen,
  Astropart.\ Phys.\  {16} (2002) 339
  [hep-ph/0009083].

\bibitem{Abazajian:2001nj}
  K.~Abazajian, G.M.~Fuller and M.~Patel,
  Phys.\ Rev.\ D {64} (2001) 023501
  [astro-ph/0101524].
  
\bibitem{Abazajian:2001vt}
  K.~Abazajian, G.M.~Fuller and W.H.~Tucker,
  Astrophys.\ J.\  {562} (2001) 593
  [astro-ph/0106002].
  
\bibitem{Abazajian:2002yz}
  K.N.~Abazajian and G.M.~Fuller,
  Phys.\ Rev.\ D {66} (2002) 023526
  [astro-ph/0204293].

\bibitem{Abazajian:2005gj}
  K.~Abazajian,
  Phys.\ Rev.\ D {73} (2006) 063506
  [astro-ph/0511630].
  
\bibitem{Abazajian:2006yn}
  K.~Abazajian and S.M.~Koushiappas,
  Phys.\ Rev.\ D {74} (2006) 023527
  [astro-ph/0605271].

\bibitem{Boyanovsky:2006it}
  D.~Boyanovsky and C.M.~Ho,
  hep-ph/0612092.

\bibitem{Boyarsky:2005us}
  A.~Boyarsky, A.~Neronov, O.~Ruchayskiy and M.~Shaposhnikov,
  Mon.\ Not.\ Roy.\ Astron.\ Soc.\  {370} (2006) 213
  [astro-ph/0512509].
  
  
\bibitem{Boyarsky:2006jm}
  A.~Boyarsky, A.~Neronov, O.~Ruchayskiy and M.~Shaposhnikov,
  JETP Lett.\  {83} (2006) 133
  [hep-ph/0601098].
  

\bibitem{Boyarsky:2006zi}
  A.~Boyarsky, A.~Neronov, O.~Ruchayskiy and M.~Shaposhnikov,
  Phys.\ Rev.\ D {74} (2006) 103506
  [astro-ph/0603368].
  
\bibitem{Asaka:2006ek}
  T.~Asaka, M.~Shaposhnikov and A.~Kusenko,
  Phys.\ Lett.\ B {638} (2006) 401
  [hep-ph/0602150].
    
  
\bibitem{Boyarsky:2006fg}
  A.~Boyarsky, A.~Neronov, O.~Ruchayskiy, M.~Shaposhnikov and I.~Tkachev,
  Phys.\ Rev. \ Lett.\ {97} (2006) 261302
  [astro-ph/0603660].
  
\bibitem{als}   
  T.~Asaka, M.~Laine and M.~Shaposhnikov,
  JHEP {06} (2006) 053
  [hep-ph/0605209].
  
  
  
\bibitem{Gelmini:2004ah}
  G.~Gelmini, S.~Palomares-Ruiz and S.~Pascoli,
  Phys.\ Rev.\ Lett.\  {93} (2004)  081302
  [astro-ph/0403323].
  
\bibitem{Riemer-Sorensen:2006fh}
  S.~Riemer-S{\o}rensen, S.H.~Hansen and K.~Pedersen,
  Astrophys.\ J.\  {644} (2006)  L33
  [astro-ph/0603661].
  
  
\bibitem{Watson:2006qb}
  C.R.~Watson, J.F.~Beacom, H.~Yuksel and T.P.~Walker,
  Phys.\ Rev.\ D {74} (2006) 033009
  [astro-ph/0605424].
  
\bibitem{Boyarsky:2006kc}
  A.~Boyarsky, O.~Ruchayskiy and M.~Markevitch,
  astro-ph/0611168.

\bibitem{Boyarsky:2006ag}
  A.~Boyarsky, J.~Nevalainen and O.~Ruchayskiy,
  astro-ph/0610961.

\bibitem{Riemer-Sorensen:2006pi}
  S.~Riemer-S{\o}rensen, K.~Pedersen, S.H.~Hansen and H.~Dahle,
  astro-ph/0610034.

\bibitem{Abazajian:2006jc}
  K.N.~Abazajian, M.~Markevitch, S.M.~Koushiappas and R.C.~Hickox,
  astro-ph/0611144.
  
\bibitem{Boyarsky:2006hr}
  A.~Boyarsky, J.-W.~den~Herder, A.~Neronov and O.~Ruchayskiy,
  astro-ph/0612219.

\bibitem{Hansen:2001zv}
  S.H.~Hansen, J.~Lesgourgues, S.~Pastor and J.~Silk,
  Mon.\ Not.\ Roy.\ Astron.\ Soc.\  {333} (2002) 544
  [astro-ph/0106108].
  
\bibitem{Viel:2005qj}
  M.~Viel, J.~Lesgourgues, M.G.~Haehnelt, S.~Matarrese and A.~Riotto,
  Phys.\ Rev.\ D {71} (2005) 063534
  [astro-ph/0501562].
  
\bibitem{Seljak:2006qw}
  U.~Seljak, A.~Makarov, P.~McDonald and H.~Trac,
  Phys.\ Rev.\ Lett.\  {97} (2006) 191303
  [astro-ph/0602430].

\bibitem{Viel:2006kd}
  M.~Viel, J.~Lesgourgues, M.G.~Haehnelt, S.~Matarrese and A.~Riotto,
   Phys.\ Rev.\ Lett.\  {97} (2006) 071301
  [astro-ph/0605706].
  
\bibitem{Tremaine:1979we}
  S.~Tremaine and J.E.~Gunn,
  Phys.\ Rev.\ Lett.\  {42} (1979) 407.

\bibitem{Lin:1983vq}
  D.N.C.~Lin and S.M.~Faber,
  Astrophys.\ J.\  {266} (1983) L21.

\bibitem{Dalcanton:2000hn}
  J.J.~Dalcanton and C.J.~Hogan,
  Astrophys.\ J.\  {561} (2001) 35
  [astro-ph/0004381].

\bibitem{Bezrukov:2006cy}
  F.~Bezrukov and M.~Shaposhnikov,
  hep-ph/0611352.
   
\bibitem{ReSigmaold}
  D.~N\"otzold and G.~Raffelt,
  Nucl.\ Phys.\ B {307} (1988) 924;
%
  K.~Enqvist, K.~Kainulainen and J.~Maalampi,
  Nucl.\ Phys.\ B {349} (1991) 754;
%
  J.C.~D'Olivo, J.F.~Nieves and M.~Torres,
  Phys.\ Rev.\ D {46} (1992) 1172.

\bibitem{ReSigma}
  C.~Quimbay and S.~Vargas-Castrill\'on,
  Nucl.\ Phys.\ B {451} (1995) 265
  [hep-ph/9504410].

\bibitem{Spergel:2006hy}
  D.N.~Spergel {\it et al.},
  astro-ph/0603449.

\bibitem{gsixg}
  K.~Kajantie, M.~Laine, K.~Rummukainen and Y.~Schr\"oder,
  Phys.\ Rev.\ D {67} (2003) 105008
  [hep-ph/0211321].
%

\bibitem{nspt}
  A.~Hietanen, K.~Kajantie, M.~Laine, K.~Rummukainen and Y.~Schr\"oder,
  JHEP {01} (2005) 013
  [hep-lat/0412008]; 
  F.~Di Renzo, M.~Laine, V.~Miccio, Y.~Schr\"oder and C.~Torrero,
  JHEP {07} (2006) 026
  [hep-ph/0605042].

\bibitem{pheneos}
  M.~Laine and Y.~Schr\"oder,
  Phys.\ Rev.\ D {73} (2006) 085009
  [hep-ph/0603048].

\bibitem{Nfcp}
  A.~Ali Khan {\it et al.}  [CP-PACS collaboration],
  Phys.\ Rev.\ D {64} (2001) 074510
  [hep-lat/0103028].

\bibitem{Nfmi}
  C.~Bernard {\it et al.},
  hep-lat/0611031.

\bibitem{Nfwu}
  Y.~Aoki, Z.~Fodor, S.D.~Katz and K.K.~Szabo,
  JHEP {01} (2006) 089
  [hep-lat/0510084].

\bibitem{Nfbi}
  S.~Ejiri, F.~Karsch, E.~Laermann and C.~Schmidt,
  Phys.\ Rev.\ D {73} (2006) 054506
  [hep-lat/0512040].

\bibitem{fp}
  P.~de Forcrand and O.~Philipsen,
  hep-lat/0607017;
%
  Y.~Aoki, G.~Endrodi, Z.~Fodor, S.D.~Katz and K.K.~Szabo,
  hep-lat/0611014.

\bibitem{Cheng:2006qk}
  M.~Cheng {\it et al.},
  Phys.\ Rev.\ D {74} (2006) 054507
  [hep-lat/0608013].


\bibitem{Aoki:2006br}
  Y.~Aoki, Z.~Fodor, S.D.~Katz and K.K.~Szabo,
  Phys.\ Lett.\ B {643} (2006) 46
  [hep-lat/0609068].

\bibitem{pdg}
  W.-M.~Yao {\it et al.}  [Particle Data Group],
  J.\ Phys.\ G {33} (2006) 1.
  
 \end{thebibliography}
\end{document}